\input harvmac
\input amssym
\input epsf


\newfam\frakfam
\font\teneufm=eufm10
\font\seveneufm=eufm7
\font\fiveeufm=eufm5
\textfont\frakfam=\teneufm
\scriptfont\frakfam=\seveneufm
\scriptscriptfont\frakfam=\fiveeufm
\def\frak{\fam\eufmfam \teneufm}


\def\bb{
\font\tenmsb=msbm10
\font\sevenmsb=msbm7
\font\fivemsb=msbm5
\textfont1=\tenmsb
\scriptfont1=\sevenmsb
\scriptscriptfont1=\fivemsb
}


\newfam\dsromfam
\font\tendsrom=dsrom10
\textfont\dsromfam=\tendsrom
\def\ds{\fam\dsromfam \tendsrom}


\newfam\mbffam
\font\tenmbf=cmmib10
\font\sevenmbf=cmmib7
\font\fivembf=cmmib5
\textfont\mbffam=\tenmbf
\scriptfont\mbffam=\sevenmbf
\scriptscriptfont\mbffam=\fivembf


\newfam\mbfcalfam
\font\tenmbfcal=cmbsy10
\font\sevenmbfcal=cmbsy7
\font\fivembfcal=cmbsy5
\textfont\mbfcalfam=\tenmbfcal
\scriptfont\mbfcalfam=\sevenmbfcal
\scriptscriptfont\mbfcalfam=\fivembfcal


\newfam\mscrfam
\font\tenmscr=rsfs10
\font\sevenmscr=rsfs7
\font\fivemscr=rsfs5
\textfont\mscrfam=\tenmscr
\scriptfont\mscrfam=\sevenmscr
\scriptscriptfont\mscrfam=\fivemscr
\def\scr{\fam\mscrfam \tenmscr}




\def\tilde{\widetilde}

\def\t{\tilde}
\def\hat{\widehat}

\def\bar{\overline}
\def\b{\bar}
\def\bsq#1{{{\b{#1}}^{\lower 2.5pt\hbox{$\scriptstyle 2$}}}}
\def\bexp#1#2{{{\b{#1}}^{\lower 2.5pt\hbox{$\scriptstyle #2$}}}}
\def\dotexp#1#2{{{#1}^{\lower 2.5pt\hbox{$\scriptstyle #2$}}}}

\def\IL{\relax{\rm I\kern-.18em L}}
\def\IH{\relax{\rm I\kern-.18em H}}
\def\IR{\relax{\rm I\kern-.18em R}}
\def\IC{\relax{\rm I\kern-0.54 em C}}

\def\rt2{\sqrt{2}}
\def\half {{1 \over 2}}
\def\Re{\mathop{\rm Re}}

\def\d{\partial}

\def\grad{\nabla}

\def\det{\mathop{\rm det}}

\def\Tr{\mathop{\rm Tr}}


\font\tenbifull=cmmib10
\font\tenbimed=cmmib7
\font\tenbismall=cmmib5
\textfont9=\tenbifull \scriptfont9=\tenbimed
\scriptscriptfont9=\tenbismall

\mathchardef\bbGamma="7000
\mathchardef\bbDelta="7001
\mathchardef\bbPhi="7002
\mathchardef\bbAlpha="7003
\mathchardef\bbXi="7004
\mathchardef\bbPi="7005
\mathchardef\bbSigma="7006
\mathchardef\bbUpsilon="7007
\mathchardef\bbTheta="7008
\mathchardef\bbPsi="7009
\mathchardef\bbOmega="700A
\mathchardef\bbalpha="710B
\mathchardef\bbbeta="710C
\mathchardef\bbgamma="710D
\mathchardef\bbdelta="710E
\mathchardef\bbepsilon="710F
\mathchardef\bbzeta="7110
\mathchardef\bbeta="7111
\mathchardef\bbtheta="7112
\mathchardef\bbiota="7113
\mathchardef\bbkappa="7114
\mathchardef\bblambda="7115
\mathchardef\bbmu="7116
\mathchardef\bbnu="7117
\mathchardef\bbxi="7118
\mathchardef\bbpi="7119
\mathchardef\bbrho="711A
\mathchardef\bbsigma="711B
\mathchardef\bbtau="711C
\mathchardef\bbupsilon="711D
\mathchardef\bbphi="711E
\mathchardef\bbchi="711F
\mathchardef\bbpsi="7120
\mathchardef\bbomega="7121
\mathchardef\bbvarepsilon="7122
\mathchardef\bbvartheta="7123
\mathchardef\bbvarpi="7124
\mathchardef\bbvarrho="7125
\mathchardef\bbvarsigma="7126
\mathchardef\bbvarphi="7127

\def\IL{\relax{\rm I\kern-.18em L}}
\def\IH{\relax{\rm I\kern-.18em H}}
\def\IR{\relax{\rm I\kern-.18em R}}
\def\IC{\relax{\rm I\kern-0.54 em C}}


\def\alphadot{{\dot\alpha}}




\def\SA{{\scr A}}
\def\SF{{\scr F}}
\def\SL{{\scr L}}
\def\CA{{\cal A}}

\def\CC{{\cal C}}
\def\CD{{\cal D}}

\def\CF{{\cal F}}

\def\CI{{\cal I}}

\def\CK{{\cal K}}
\def\CL{{\cal L}}
\def\CM{{\cal M}}
\def\CN{{\cal N}}
\def\CO{{\cal O}}

\def\CS{{\cal S}}

\def\CV{{\cal V}}

\def\CX{{\cal X}}
\def\CY{{\cal Y}}
\def\CZ{{\cal Z}}


\def\1{{\ds 1}}
\def\R{\hbox{$\bb R$}}
\def\C{\hbox{$\bb C$}}

\def\Z{\hbox{$\bb Z$}}


\def\ep{\varepsilon}

\def\fr{{ r} 
}

\noblackbox

\def\unit{\relax{\rm 1\kern-.26em I}}
\def\nada{\relax{\rm 0\kern-.30em l}}
\def\tilde{\widetilde}
\def\t{\tilde}
\def\alphadot{{\dot \alpha}}

\def\det{{\rm det}}

\noblackbox
\def\IL{\relax{\rm I\kern-.18em L}}
\def\IH{\relax{\rm I\kern-.18em H}}
\def\IR{\relax{\rm I\kern-.18em R}}
\def\IC{\relax\hbox{$\inbar\kern-.3em{\rm C}$}}
\def\IZ{\relax\ifmmode\mathchoice
{\hbox{\cmss Z\kern-.4em Z}}{\hbox{\cmss Z\kern-.4em Z}} {\lower.9pt\hbox{\cmsss Z\kern-.4em Z}}
{\lower1.2pt\hbox{\cmsss Z\kern-.4em Z}}\else{\cmss Z\kern-.4em Z}\fi}
\def\CM {{\cal M}}
\def\CN {{\cal N}}

\def\CD {{\cal D}}
\def\CF {{\cal F}}

\def\partialslash{\not{\hbox{\kern-2pt $\partial$}}}

\def\CL {{\cal L}}
\def\CV {{\cal V}}
\def\CO {{\cal O}}
\def\CZ {{\cal Z}}

\def\CC {{\cal C}}

\def\CS {{\cal S}}
\def\CA{{\cal A}}
\def\CK{{\cal K}}
\def\CM {{\cal M}}
\def\CN {{\cal N}}

\def\CO {{\cal O}}

\def\CV{{\cal V }}

\def\CZ {{\cal Z }}
\def\CS {{\cal S }}

\def\CY{{\cal Y }}

\def\det{{\rm det}}
\def\Tr{{\rm Tr}}

\font\manual=manfnt \def\dbend{\lower3.5pt\hbox{\manual\char127}}

\def\IZ{\relax\ifmmode\mathchoice
{\hbox{\cmss Z\kern-.4em Z}}{\hbox{\cmss Z\kern-.4em Z}} {\lower.9pt\hbox{\cmsss Z\kern-.4em Z}}
{\lower1.2pt\hbox{\cmsss Z\kern-.4em Z}}\else{\cmss Z\kern-.4em Z}\fi}
\def\half {{1\over 2}}

\def\bar{\overline}
\def\CS{{\cal S}}

\def\rt2{\sqrt{2}}
\def\irt2{{1\over\sqrt{2}}}

\def\t{\tilde}
\def\hat{\widehat}
\def\slashchar#1{\setbox0=\hbox{$#1$}           
   \dimen0=\wd0                                 
   \setbox1=\hbox{/} \dimen1=\wd1               
   \ifdim\dimen0>\dimen1                        
      \rlap{\hbox to \dimen0{\hfil/\hfil}}      
      #1                                        
   \else                                        
      \rlap{\hbox to \dimen1{\hfil$#1$\hfil}}   
      /                                         
   \fi}

\def\foursqr#1#2{{\vcenter{\vbox{
    \hrule height.#2pt
    \hbox{\vrule width.#2pt height#1pt \kern#1pt
    \vrule width.#2pt}
    \hrule height.#2pt
    \hrule height.#2pt
    \hbox{\vrule width.#2pt height#1pt \kern#1pt
    \vrule width.#2pt}
    \hrule height.#2pt
        \hrule height.#2pt
    \hbox{\vrule width.#2pt height#1pt \kern#1pt
    \vrule width.#2pt}
    \hrule height.#2pt
        \hrule height.#2pt
    \hbox{\vrule width.#2pt height#1pt \kern#1pt
    \vrule width.#2pt}
    \hrule height.#2pt}}}}
\def\psqr#1#2{{\vcenter{\vbox{\hrule height.#2pt
    \hbox{\vrule width.#2pt height#1pt \kern#1pt
    \vrule width.#2pt}
    \hrule height.#2pt \hrule height.#2pt
    \hbox{\vrule width.#2pt height#1pt \kern#1pt
    \vrule width.#2pt}
    \hrule height.#2pt}}}}
\def\sqr#1#2{{\vcenter{\vbox{\hrule height.#2pt
    \hbox{\vrule width.#2pt height#1pt \kern#1pt
    \vrule width.#2pt}
    \hrule height.#2pt}}}}

\def\figin{\epsfcheck\figin}\def\figins{\epsfcheck\figins}
\def\epsfcheck{\ifx\epsfbox\UnDeFiNeD
\message{(NO epsf.tex, FIGURES WILL BE IGNORED)}
\gdef\figin##1{\vskip2in}\gdef\figins##1{\hskip.5in}
\else\message{(FIGURES WILL BE INCLUDED)}%
\gdef\figin##1{##1}\gdef\figins##1{##1}\fi}
\def\DefWarn#1{}
\def\figinsert{\goodbreak\midinsert}
\def\ifig#1#2#3{\DefWarn#1\xdef#1{fig.~\the\figno}
\writedef{#1\leftbracket fig.\noexpand~\the\figno}%
\figinsert\figin{\centerline{#3}}\medskip\centerline{\vbox{\baselineskip12pt \advance\hsize by
-1truein\noindent\footnotefont{\bf Fig.~\the\figno:\ } \it#2}}
\bigskip\endinsert\global\advance\figno by1}


\lref\FestucciaWS{
  G.~Festuccia and N.~Seiberg,
  ``Rigid Supersymmetric Theories in Curved Superspace,''
JHEP {\bf 1106}, 114 (2011).
[arXiv:1105.0689 [hep-th]].
}

\lref\MartelliFU{
  D.~Martelli, A.~Passias and J.~Sparks,
  ``The Gravity dual of supersymmetric gauge theories on a squashed three-sphere,''
Nucl.\ Phys.\ B {\bf 864}, 840 (2012).
[arXiv:1110.6400 [hep-th]].
}

\lref\ClossetVG{
  C.~Closset, T.~T.~Dumitrescu, G.~Festuccia, Z.~Komargodski and N.~Seiberg,
  ``Contact Terms, Unitarity, and F-Maximization in Three-Dimensional Superconformal Theories,''
JHEP {\bf 1210}, 053 (2012).
[arXiv:1205.4142 [hep-th]].
}

\lref\ClossetVP{
  C.~Closset, T.~T.~Dumitrescu, G.~Festuccia, Z.~Komargodski and N.~Seiberg,
  ``Comments on Chern-Simons Contact Terms in Three Dimensions,''
JHEP {\bf 1209}, 091 (2012).
[arXiv:1206.5218 [hep-th]].
}

\lref\KomargodskiRB{
  Z.~Komargodski and N.~Seiberg,
  ``Comments on Supercurrent Multiplets, Supersymmetric Field Theories and Supergravity,''
JHEP {\bf 1007}, 017 (2010).
[arXiv:1002.2228 [hep-th]].
}

\lref\SohniusTP{
  M.~F.~Sohnius and P.~C.~West,
  ``An Alternative Minimal Off-Shell Version of N=1 Supergravity,''
Phys.\ Lett.\ B {\bf 105}, 353 (1981).
}

\lref\DumitrescuHA{
  T.~T.~Dumitrescu, G.~Festuccia and N.~Seiberg,
  ``Exploring Curved Superspace,''
JHEP {\bf 1208}, 141 (2012).
[arXiv:1205.1115 [hep-th]].
}

\lref\DumitrescuAT{
  T.~T.~Dumitrescu and G.~Festuccia,
  ``Exploring Curved Superspace (II),''
JHEP {\bf 1301}, 072 (2013).
[arXiv:1209.5408 [hep-th]].
}

\lref\DumitrescuIU{
  T.~T.~Dumitrescu and N.~Seiberg,
  ``Supercurrents and Brane Currents in Diverse Dimensions,''
JHEP {\bf 1107}, 095 (2011).
[arXiv:1106.0031 [hep-th]].
}

\lref\SohniusFW{
  M.~Sohnius and P.~C.~West,
  ``The Tensor Calculus And Matter Coupling Of The Alternative Minimal Auxiliary Field Formulation Of N=1 Supergravity,''
Nucl.\ Phys.\ B {\bf 198}, 493 (1982).
}

\lref\WittenEV{
  E.~Witten,
  ``Supersymmetric Yang-Mills theory on a four manifold,''
J.\ Math.\ Phys.\  {\bf 35}, 5101 (1994).
[hep-th/9403195].
}

\lref\KlareGN{
  C.~Klare, A.~Tomasiello and A.~Zaffaroni,
  ``Supersymmetry on Curved Spaces and Holography,''
JHEP {\bf 1208}, 061 (2012).
[arXiv:1205.1062 [hep-th]].
}

\lref\RomelsbergerEG{
  C.~Romelsberger,
  ``Counting chiral primaries in N = 1, d=4 superconformal field theories,''
Nucl.\ Phys.\ B {\bf 747}, 329 (2006).
[hep-th/0510060].
}

\lref\ClossetRU{
  C.~Closset, T.~T.~Dumitrescu, G.~Festuccia and Z.~Komargodski,
  ``Supersymmetric Field Theories on Three-Manifolds,''
JHEP {\bf 1305}, 017 (2013).
[arXiv:1212.3388].
}

\lref\Kobayashi{
S.~Kobayashi, ``Differential Geometry of Complex Vector Bundles,'' Princeton University Press (1987).
}

\lref\KinneyEJ{
  J.~Kinney, J.~M.~Maldacena, S.~Minwalla and S.~Raju,
  ``An Index for 4 dimensional super conformal theories,''
Commun.\ Math.\ Phys.\  {\bf 275}, 209 (2007).
[hep-th/0510251].
}

\lref\CassaniDBA{
  D.~Cassani and D.~Martelli,
  ``Supersymmetry on curved spaces and superconformal anomalies,''
JHEP {\bf 1310}, 025 (2013).
[arXiv:1307.6567 [hep-th]].
}

\lref\JohansenAW{
  A.~Johansen,
  ``Twisting of $N=1$ SUSY gauge theories and heterotic topological theories,''
Int.\ J.\ Mod.\ Phys.\ A {\bf 10}, 4325 (1995).
[hep-th/9403017].
}

\lref\KapustinKZ{
  A.~Kapustin, B.~Willett and I.~Yaakov,
  ``Exact Results for Wilson Loops in Superconformal Chern-Simons Theories with
  Matter,''
  JHEP {\bf 1003}, 089 (2010)
  [arXiv:0909.4559 [hep-th]].
}

\lref\JafferisUN{
  D.~L.~Jafferis,
  ``The Exact Superconformal R-Symmetry Extremizes Z,''
JHEP {\bf 1205}, 159 (2012).
[arXiv:1012.3210 [hep-th]].
}

\lref\HamaAV{
  N.~Hama, K.~Hosomichi and S.~Lee,
  ``Notes on SUSY Gauge Theories on Three-Sphere,''
JHEP {\bf 1103}, 127 (2011).
[arXiv:1012.3512 [hep-th]].
}

\lref\HamaEA{
  N.~Hama, K.~Hosomichi and S.~Lee,
  ``SUSY Gauge Theories on Squashed Three-Spheres,''
JHEP {\bf 1105}, 014 (2011).
[arXiv:1102.4716 [hep-th]].
}

\lref\ClossetVRA{
  C.~Closset, T.~T.~Dumitrescu, G.~Festuccia and Z.~Komargodski,
  ``The Geometry of Supersymmetric Partition Functions,''
JHEP {\bf 1401}, 124 (2014).
[arXiv:1309.5876 [hep-th]].
}

\lref\AsselPAA{
  B.~Assel, D.~Cassani and D.~Martelli,
JHEP {\bf 1408}, 123 (2014).
[arXiv:1405.5144 [hep-th]].
}

\lref\WittenZE{
  E.~Witten,
  ``Topological Quantum Field Theory,''
Commun.\ Math.\ Phys.\  {\bf 117}, 353 (1988).
}

\lref\BeniniMF{
  F.~Benini, C.~Closset and S.~Cremonesi,
  ``Comments on 3d Seiberg-like dualities,''
JHEP {\bf 1110}, 075 (2011).
[arXiv:1108.5373 [hep-th]].
}

\lref\FerraraQXA{
  S.~Ferrara and S.~Sabharwal,
  ``Structure of New Minimal Supergravity,''
Annals Phys.\  {\bf 189}, 318 (1989).
}

\lref\vyasphd{
K.~Vyas,
  ``Topics in topological and holomorphic quantum field theory,'' Caltech Ph.D. Thesis (2010), [http://resolver.caltech.edu/CaltechTHESIS:06012010-010858409].
}

\lref\IntriligatorJJ{
  K.~A.~Intriligator and B.~Wecht,
  ``The Exact superconformal R symmetry maximizes a,''
Nucl.\ Phys.\ B {\bf 667}, 183 (2003).
[hep-th/0304128].
}

\lref\ClossetSXA{
  C.~Closset and I.~Shamir,
  ``The $\CN=1$ Chiral Multiplet on $T^2\times S^2$ and Supersymmetric Localization,''
JHEP {\bf 1403}, 040 (2014).
[arXiv:1311.2430 [hep-th]].
}

\lref\LawsonYR{
  H.~B.~Lawson and M.~L.~Michelsohn,
  ``Spin geometry,'' Princeton mathematical series, Vol.~38 (1989).
}

\lref\BG{
M.~Brunella and~E.~Ghys, ``Umbilical Foliations and Transversely Holomorphic Flows,'' J.~Differential Geometry {\bf 41} 1 (1995).
}

\lref\DKi{
T.~Duchamp and~M.~Kalka, ``Deformation Theory for Holomorphic Foliations,'' J.~Differential Geometry {\bf 14} 317 (1979).
}

\lref\GM{
X.~Gomez-Mont, ``Transversal Holomorphic Structures,'' J.~Differential Geometry {\bf 15} 161 (1980).
}

\lref\Haef{
J.~Girbau, A.~Haefliger, and~D.~Sundararaman, ``On deformations of transversely holomorphic foliations,'' Journal f\"ur die reine und angewandte Mathematik {\bf 345} 122 (1983).
}

\lref\Ghys{
E.~Ghys, ``On transversely holomorphic flows II,'' Inventiones mathematicae {\bf 126} 281 (1996).
}

\lref\Brunella{
M.~Brunella, ``On transversely holomorphic flows I,'' Inventiones mathematicae {\bf 126} 265 (1996).
}

\lref\KarlhedeAX{
  A.~Karlhede and M.~Rocek,
  ``Topological Quantum Field Theory and $N=2$ Conformal Supergravity,''
Phys.\ Lett.\ B {\bf 212}, 51 (1988).
}

\lref\NekrasovQD{
  N.~A.~Nekrasov,
  ``Seiberg-Witten prepotential from instanton counting,''
Adv.\ Theor.\ Math.\ Phys.\  {\bf 7}, 831 (2003).
[hep-th/0206161].
}

\lref\NekrasovRJ{
  N.~Nekrasov and A.~Okounkov,
  ``Seiberg-Witten theory and random partitions,''
[hep-th/0306238].
}

\lref\PestunRZ{
  V.~Pestun,
  ``Localization of gauge theory on a four-sphere and supersymmetric Wilson loops,''
Commun.\ Math.\ Phys.\  {\bf 313}, 71 (2012).
[arXiv:0712.2824 [hep-th]].
}

\lref\KlareDKA{
  C.~Klare and A.~Zaffaroni,
  ``Extended Supersymmetry on Curved Spaces,''
JHEP {\bf 1310}, 218 (2013).
[arXiv:1308.1102 [hep-th]].
}

\lref\AharonyDHA{
  O.~Aharony, S.~S.~Razamat, N.~Seiberg and B.~Willett,
  ``3d dualities from 4d dualities,''
JHEP {\bf 1307}, 149 (2013).
[arXiv:1305.3924 [hep-th]].
}

\lref\KuzenkoUYA{
  S.~M.~Kuzenko, U.~Lindstrom, M.~Rocek, I.~Sachs and G.~Tartaglino-Mazzucchelli,
  ``Three-dimensional N=2 supergravity theories: From superspace to components,''
Phys.\ Rev.\ D {\bf 89}, 085028 (2014).
[arXiv:1312.4267 [hep-th]].
}



\rightline{WIS/05/14-JUN-DPPA}
\vskip-20pt
\Title{
} {\vbox{\centerline{From Rigid Supersymmetry}
\centerline{to Twisted Holomorphic Theories}}}
\vskip-15pt
\centerline{Cyril Closset,$^1$ Thomas T. Dumitrescu,$^{2}$ Guido Festuccia,$^3$ Zohar Komargodski$^{4}$}
\vskip15pt
\centerline{ $^{1}$ {\it Simons Center for Geometry and Physics,}}
  \vskip-5pt
  \centerline{\it Stony Brook University, Stony Brook, NY 11794, USA}
 \centerline{$^{2}$ {\it Department of Physics, Harvard University, Cambridge, MA 02138, USA}}
 \centerline{$^{3}${\it Niels Bohr International Academy and Discovery Center, Niels Bohr Institute,}} 
  \vskip-5pt
  \centerline{\it University of Copenhagen, Blegdamsvej 17, 2100 Copenhagen~\O, Denmark}
  \centerline{ $^{4}$ {\it Weizmann Institute of Science, Rehovot
76100, Israel}}

\vskip30pt

\noindent We study~$\CN=1$ field theories with a~$U(1)_R$ symmetry on compact four-manifolds~$\CM$. Supersymmetry requires~$\CM$ to be a complex manifold. The supersymmetric theory on~$\CM$ can be described in terms of conventional fields coupled to background supergravity, or in terms of twisted fields adapted to the complex geometry of~$\CM$. Many properties of the theory that are difficult to see in one formulation are simpler in the other one. We use the twisted description to study the dependence of the partition function~$Z_\CM$ on the geometry of~$\CM$, as well as coupling constants and background gauge fields, recovering and extending previous results. We also indicate how to generalize our analysis to three-dimensional $\CN=2$ theories with a~$U(1)_R$ symmetry. In this case supersymmetry requires~$\CM$ to carry a transversely holomorphic foliation, which endows it with a near-perfect analogue of complex geometry. Finally, we present new explicit formulas for the dependence of~$Z_\CM$ on the choice of~$U(1)_R$ symmetry in four and three dimensions, and illustrate them for complex manifolds diffeomorphic to~$S^3 \times S^1$, as well as general squashed three-spheres.

\Date{July 2014}

\listtoc \writetoc

\newsec{Introduction}

Broadly speaking, there are two systematic ways to construct supersymmetric field theories on a Riemannian manifold~$\CM$. (For simplicity, we will assume that~$\CM$ is compact.) 
\smallskip
{\it Twisting:} In this approach, a subgroup of Euclidean spacetime rotations is identified with the~$R$-symmetry group, so that some of the supercharges transform as scalars and can be defined globally on~$\CM$. In curved space, the role of spacetime rotations is played by the Riemannian holonomy group. The archetypal (and also the first) example of this procedure is~$\CN=2$ Yang-Mills theory twisted on a four-manifold~\WittenZE, where an~$SU(2)$ factor of the~$SU(2)_\ell \times SU(2)_r$ holonomy group is identified with the~$SU(2)_R$ symmetry. After twisting, the bosonic and fermionic fields of the theory become differential forms on~$\CM$, and it can be shown that the partition function~$Z_\CM$ is independent of the metric. For this reason, the twist is referred to as topological. Four-dimensional~$\CN=1$ theories with a~$U(1)_R$ symmetry can be twisted on K\"ahler manifolds, which have~$U(2)$ holonomy~\refs{\JohansenAW,\WittenEV}. In this case the twist is not topological, because it depends on a choice of K\"ahler structure.  
\smallskip
{\it Rigid Supersymmetry:} In this approach, supersymmetric theories on~$\CM$ are described in terms of conventional field variables coupled to background off-shell supergravity~\FestucciaWS. The allowed supersymmetric configurations for the bosonic supergravity fields are determined by setting the fermions in the supergravity multiplet and their supersymmetry variations to zero. These constraints often imply the presence of additional geometric structures on~$\CM$ (other than the Riemannian metric~$g_{\mu\nu}$), which affect the supersymmetry transformations of dynamical fields, as well as the Lagrangian on~$\CM$.
\smallskip
We will explore the relationship between twisting and rigid supersymmetry for~$\CN=1$ theories with a~$U(1)_R$ symmetry in four dimensions, and for their three-dimensional cousins.\foot{In the context of four-dimensional~$\CN=2$ theories, the relationship between the topologically twisted formulation and a description in terms of conventional field variables was discussed in~\refs{\KarlhedeAX\NekrasovQD\NekrasovRJ\PestunRZ-\KlareDKA}.} The latter approach is based on the new-minimal supergravity multiplet constructed in~\refs{\SohniusTP,\SohniusFW}, which contains the metric~$g_{\mu\nu}$, an~$R$-symmetry gauge field~$A_\mu^{(R)}$, and a covariantly conserved vector~$V^\mu$, as well as the gravitinos~$\Psi_{\mu\alpha}, \t \Psi_{\mu\alphadot}$.\foot{Unless stated otherwise, we follow the conventions of~\ClossetVRA.} Setting the gravitinos and their supersymmetry variations to zero leads to the following generalized Killing spinor equations~\FestucciaWS,
\eqn\speq{\eqalign{& \big(\grad_\mu - i A^{( R)}_\mu\big) \zeta = - {i\over 2} V^\nu \sigma_{\mu}\t \sigma_{\nu} \zeta~,\cr
&\big(\grad_\mu + i A^{( R)}_\mu\big) \t \zeta = { i\over 2} V^\nu \t \sigma_{\mu}\sigma_{\nu} \t \zeta~,}}
where~$\zeta$ and~$\t \zeta$ carry~$R$-charges~$+1$ and~$-1$. The existence of such Killing spinors implies that the bosonic background fields preserve a supercharge~$Q$ or~$\t Q$ of~$R$-charge~$-1$ or~$+1$, respectively. We will mostly focus on manifolds~$\CM$ that admit a Killing spinor~$\zeta_\alpha$ of~$R$-charge~$+1$, which exists if and only if~$\CM$ possess an integrable complex structure~${J^\mu}_\nu$ and~$g_{\mu\nu}$ is a compatible Hermitian metric~\refs{\DumitrescuHA,\KlareGN}. The other supergravity fields~$A_\mu^{(R)}$ and~$V^\mu$ are then essentially determined in terms of~${J^\mu}_\nu$ and~$g_{\mu\nu}$. 

The rigid supersymmetry approach cleanly separates Lagrangians and transformation rules, which only depend on the supergravity fields, from the allowed supersymmetric backgrounds, which are determined by solving the Killing spinor equations~\speq. These backgrounds depend on the Hermitian structure, and hence the same is true for the transformation rules and the Lagrangian on~$\CM$. However, attempting to express the latter directly in terms of~${J^\mu}_\nu$ and~$g_{\mu\nu}$ quickly leads to complicated formulas. This makes it challenging to study the dependence of supersymmetric observables, such as the  partition function~$Z_\CM$, on the geometry of~$\CM$. 

This problem was sidestepped in~\ClossetVRA\ by first analyzing the dependence of~$Z_\CM$ on~${J^\mu}_\nu$ and~$g_{\mu\nu}$ at the linearized level around flat space and arguing for the validity of the results on a general manifold~$\CM$. (The second step relies on the global properties of the supercharge~$Q$ on~$\CM$, which will be reviewed below.) At the linearized level, the Lagrangians on~$\CM$ and in flat space only differ by operators that reside in the~$R$-current supermultiplet of the flat-space theory~\refs{\KomargodskiRB,\DumitrescuIU}, which also contains the supersymmetry current and the energy-momentum tensor. The sources that multiply these operators are the linearized new-minimal supergravity fields, which in turn depend on~${J^\mu}_\nu$ and~$g_{\mu\nu}$. Using the known transformation rules of the~$R$-current supermultiplet and the fact that~$Q$-exact terms in the Lagrangian do not affect the partition function~$Z_\CM$, it was shown that~$Z_\CM$ is a locally holomorphic function of the complex structure moduli (finitely many, since~$\CM$ is compact), but independent of the Hermitian metric. Similarly, it was found that~$Z_\CM$ only depends on background gauge fields for Abelian flavor symmetries, which determine holomorphic line bundles over~$\CM$, through a locally holomorphic function of the corresponding bundle moduli (again, finitely many). 

The arguments in~\ClossetVRA\ only rely on general properties of the~$R$-current supermultiplet. Therefore, they are model independent and do not require a Lagrangian description. In sections~5 and 6 we will use similar methods to study the dependence of the partition function on the choice of~$R$-symmetry in four and three dimensions. An obvious shortcoming of the linearized approach is the need to explain why results obtained in a neighborhood of flat space continue to hold for arbitrary~$\CM$. Moreover, this approach inherits the emphasis of the rigid supersymmetry paradigm on supergravity fields. The Hermitian structure, which underlies many aspects of the supersymmetric theory on~$\CM$, only enters indirectly through the dependence of the supergravity background fields on~${J^\mu}_\nu$ and~$g_{\mu\nu}$. 

Here we will pursue a more geometric, fully non-linear approach, which harnesses the complex geometry of~$\CM$. As was shown in~\DumitrescuHA, the properties of the Killing spinor~$\zeta_\alpha$ allow us to identify~$L = \CK^{-\half}$, where~$L$ is the~$U(1)_R$ bundle and~$\CK$ is the canonical line bundle of complex~$(2,0)$-forms on~$\CM$. This defines a twist, which turns fields of~$R$-charge~$r$ into sections of~$\CK^{-{r \over 2}}$. In particular, the supercharge~$Q$ corresponding to~$\zeta_\alpha$ transforms as a scalar under holomorphic coordinate changes.  This allows us to make contact with the twisted~$\CN=1$ theories on~K\"ahler manifolds studied in~\refs{\JohansenAW,\WittenEV,\vyasphd}, and to generalize them to arbitrary complex manifolds~$\CM$. Since the twist is defined with respect to a fixed complex structure on~$\CM$, we will follow~\vyasphd\ and refer to it as a holomorphic twist. An example of recent interest is furnished by complex manifolds diffeomorphic to~$S^3 \times S^1$, known as primary Hopf surfaces, which are not K\"ahler. Partition functions on these spaces compute the supersymmetric indices defined in~\refs{\RomelsbergerEG,\KinneyEJ}, up to local counterterms and effects due to quantum anomalies~\refs{\FestucciaWS,\ClossetVRA,\ClossetSXA,\AsselPAA}. 

In this paper, we will develop the twisted holomorphic formulation for~$\CN=1$ theories on complex manifolds~$\CM$, and use it to study the properties of their partition functions~$Z_\CM$. This will allow us to give an alternative, more direct, proof of the results in~\ClossetVRA\ about the dependence of~$Z_\CM$ on the geometry of~$\CM$, and to extend them by obtaining exact statements about the dependence of~$Z_\CM$ on other continuous parameters. For instance, we will show that~$Z_\CM$ is a locally holomorphic function of anti-chiral~$F$-term coupling constants, but that it does not depend on~$D$-terms or chiral~$F$-terms. We will also study the dependence of~$Z_\CM$ on continuous shifts of the~$R$-symmetry by an Abelian flavor symmetry, which is completely fixed by its holomorphic dependence on complex structure and line bundle moduli. At the end of the paper, we will sketch how to generalize our results to three-dimensional~$\CN=2$ theories with a~$U(1)_R$ symmetry (see the outline below). 

As was already explained in~\ClossetVRA, our methods are essentially classical: we will assume that~$Q$-exact terms in the Lagrangian do not affect the partition function~$Z_\CM$, and that it is legitimate to perform field redefinitions. Both assumptions can be ruined by quantum effects, hidden in the path integral measure, and hence our results about the partition function~$Z_\CM$ only hold up to local counterterms and possible quantum anomalies. This qualification applies throughout the paper, although we will repeat it occasionally. In this context, some effects due to anomalies were recently discussed in~\refs{\CassaniDBA,\AsselPAA}. 

The remainder of this paper is structured as follows: in section~2, we review rigid supersymmetric theories on four-manifolds~$\CM$ that admit solutions to the Killing spinor equations~\speq. In particular, we assemble the ingredients needed to describe renormalizable theories of chiral multiplets and gauge fields coupled to background supergravity. 

Section~3 recalls the correspondence between a Killing spinor~$\zeta_\alpha$ and a Hermitian structure on~$\CM$, and how this defines the holomorphic twist. We exhibit an explicit field redefinition from conventional to twisted field variables. In terms of the twisted variables, which are adapted to the complex geometry of~$\CM$, the supersymmetry transformations and Lagrangians described in section~2 simplify dramatically. This allows us to prove that~$D$-terms and chiral~$F$-terms are~$Q$-exact, and hence the partition function~$Z_\CM$ does not depend on 
coupling constants that multiply such terms. 

In section~4, we use twisted variables to study the dependence of~$Z_\CM$ on the complex structure and Hermitian metric on~$\CM$, as well as Abelian background gauge fields that couple to flavor symmetries, which define holomorphic line bundles over~$\CM$. We recover the results of~\ClossetVRA, already summarized above: the partition function is a locally holomorphic function of finitely many complex structure and holomorphic line bundle moduli; it does not depend on the Hermitian metric or the detailed configuration of background flavor gauge fields. The proof using twisted variables is manifestly non-linear and fully utilizes the complex geometry of~$\CM$. However, unlike the model-independent analysis in~\ClossetVRA, it requires an explicit Lagrangian description in terms of fields. 

Section~5 examines the dependence of the partition function on continuous shifts of the~$R$-symmetry by an Abelian flavor symmetry, which are possible if the canonical bundle of~$\CM$ is topologically trivial. We derive an explicit formula that expresses the~$R$-symmetry dependence of~$Z_\CM$ in terms of its dependence on complex structure and holomorphic line bundle moduli, and illustrate this formula for primary Hopf surfaces, i.e.~complex manifolds diffeomorphic to~$S^3 \times S^1$.

In section~6, we explain how to generalize the results of this paper to three-dimensional~$\CN=2$ theories with a~$U(1)_R$ symmetry on curved manifolds~$\CM$. Supersymmetry requires~$\CM$ to possess a transversely holomorphic foliation~\refs{\ClossetRU,\ClossetVRA}, which endows it with a near-perfect analogue of complex geometry. Therefore, supersymmetric theories on~$\CM$ also have a twisted description, which can be used to re-derive and extend the results of~\ClossetVRA\ on the parameter dependence of the three-dimensional partition function~$Z_\CM$. We generalize the results of section~5 to obtain an explicit formula for the~$R$-symmetry dependence of~$Z_\CM$, and we apply it to a general squashed sphere. Our formula provides an {\it a priori} explanation for the `mysterious' holomorphy observed in~\JafferisUN\ on a round sphere, and its generalization to the squashed case.

\newsec{Supersymmetry Multiplets and Lagrangians}

In this section, we review the rigid supersymmetry algebra and its multiplets on manifolds~$\CM$ that admit one or several solutions to the Killing spinor equations~\speq. Using these multiplets, it is straightforward to write down supersymmetric Lagrangians on~$\CM$. We will give explicit formulas for renormalizable theories of chiral multiplets coupled to dynamical or background gauge fields. As explained in~\FestucciaWS, these results follow from the corresponding formulas in matter-coupled new minimal supergravity~\refs{\SohniusTP,\SohniusFW}. The rigid curved-space Lagrangians obtained in this way reduce to conventional flat-space Lagrangians at short distances. In particular, taking the metric to be flat and setting to zero all other background fields results in a theory with full~$\CN=1$ super-Poincar\'e invariance. An example of a supersymmetric term that does not seem to arise from a known supergravity-matter coupling will be discussed in section~3. 

\subsec{The Supersymmetry Algebra}

Given solutions~$\zeta, \eta$ and~$\t \zeta, \t \eta$ of the Killing spinor equations~\speq, the rigid supersymmetry algebra on~$\CM$ follows from the algebra of local supersymmetry transformations in new minimal supergavity~\refs{\SohniusTP,\SohniusFW}. If~$\Phi^{(r)}$ is a field of~$R$-charge~$r$ and arbitrary spin,
\eqn\susyalg{\eqalign{
&\{\delta_\zeta, \delta_{\t \zeta}\}\Phi^{(r)} = 2 i \CL'_K \Phi^{(r)}~, \qquad K^\mu = \zeta \sigma^\mu \t \zeta~,\cr
& \{\delta_\zeta, \delta_\eta\} \Phi^{(r)} = \{\delta_{\t \zeta}, \delta_{\t \eta}\} \Phi^{(r)} = 0~.}}
Here we have assumed that~$\Phi^{(r)}$ does not couple to other dynamical or background gauge fields, which will be discussed below. The Killing spinor equations~\speq\ imply that~$K^\mu$ is a Killing vector~\DumitrescuHA. The modified Lie derivative~$\CL'_K$ is covariant with respect to local~$R$-symmetry gauge transformations, 
\eqn\modlie{
\CL'_K \Phi^{(r)} = \CL_K \Phi^{(r)} - i r K^\mu \Big(A_\mu^{(R)} + {3 \over 2}V_\mu\Big)\Phi^{(r)}~,
}
where~$\CL_K$ is the usual Lie derivative. Two additional comments are in order:
\medskip
\item{1.)} We take the Killing spinors parametrizing supersymmetry transformations to be commuting. Consequently, supersymmetry transformations~$\delta$ are anti-commuting. This is reflected in the structure of the algebra~\susyalg.
\smallskip
\item{2.)} Since we are working in Euclidean signature, left-handed and right-handed spinors are not related by complex conjugation. Compatibility with supersymmetry transformations prompts us to formally relax the reality conditions on bosonic fields as well. We therefore view all fields as independent complex variables, with the understanding that one must ultimately choose an integration contour for dynamical bosonic fields.

\subsec{Basic Supersymmetry Multiplets}

We will now discuss the realization of the algebra~\susyalg\ on a general multiplet~$S$, whose bottom component~$C$ is a complex scalar of $R$-charge~$r$. Although there are more general multiplets, whose bottom components carry spin, we will obtain all necessary multiplets by starting with~$S$ and imposing supersymmetric constraints. 

The general multiplet~$S$ has~$16+16$ components,
\eqn\genSfourd{
S= (C, \chi_\alpha, \t \chi^\alphadot, M, \t M, a_\mu, \lambda_\alpha, \t \lambda^\alphadot, D)~,
}
whose~$R$-charges relative to its bottom component are given by~$\left(0,-1,1,-2,2,0,1,-1,0\right)$. Its transformation rules under a supersymmetry variation with spinor parameters~$\zeta, \t \zeta$ can be obtained by taking a rigid limit of the tensor calculus for new-minimal supergravity developed in~\refs{\SohniusFW, \FerraraQXA},
\eqn\fdgmul{\eqalign{& \delta C = i \zeta \chi - i \t \zeta \t \chi~,\cr
& \delta \chi = \zeta M + \sigma^\mu \t \zeta \, (i a_\mu + D_\mu C)~, \qquad \cr
& \delta \t \chi = \t \zeta \t M + \t \sigma^\mu \zeta\, (i a_\mu - D_\mu C)~,\cr
& \delta M = 2 \t \zeta \t \lambda + 2 i D_\mu (\t \zeta \, \t \sigma^\mu \chi)~, \qquad \cr
& \delta \t M = 2 \zeta \lambda + 2 i D_\mu (\zeta \sigma^\mu \t \chi)~,\cr
& \delta a_\mu = i (\zeta \sigma_\mu \t \lambda + \t \zeta\, \t \sigma_\mu \lambda) + D_\mu (\zeta \chi + \t \zeta \t \chi)~,\cr
& \delta \lambda = i \zeta D + 2 \sigma^{\mu\nu}\zeta \, D_\mu a_\nu~, \qquad \cr
& \delta \t \lambda = - i \t \zeta D + 2 \t \sigma^{\mu\nu} \t \zeta \, D_\mu a_\nu~,\cr
& \delta D = - D_\mu (\zeta \sigma^\mu \t \lambda - \t \zeta \, \t \sigma^\mu \lambda) + 2 i V_\mu (\zeta \sigma^\mu \t \lambda + \t \zeta \, \t \sigma^\mu \lambda) \cr
&\hskip25pt  + {i \fr \over 4} (R - 6 V^\mu V_\mu) (\zeta \chi + \t \zeta \t \chi)~.}}
Here~$R$ is the Ricci scalar. The action of the covariant derivative~$D_\mu$ on a field~$\Phi^{(r)}$ of~$R$-charge~$r$ is given by
\eqn\covddef{
D_\mu \Phi^{(r)} = \Big(\grad_\mu - i r \big(A^{(R)}_\mu+{3\over 2} V_\mu\big)\Big) \Phi^{(r)}~,
}
with~$\grad_\mu$ the usual Levi-Civita connection. It can be checked that the transformations in~\fdgmul\ realize the algebra~\susyalg\ whenever the Killing spinors satisfy~\speq. 

Given two general multiplets~$S_1, S_2$ with bottom components~$C_1, C_2$ of~$R$-charge~$\fr_1, \fr_2$, we can define a product multiplet~$S$, whose bottom component~$C = C_1 C_2$ carries~$R$-charge~$\fr_1 + \fr_2$. The other components of~$S$ can then be expressed in terms of the components of~$S_1, S_2$ by repeatedly applying the transformations in~\fdgmul. The resulting multiplication rules are summarized in~appendix A.

In order to construct supersymmetric Lagrangians, we need chiral and anti-chiral matter multiplets, as well as vector multiplets containing gauge fields. As in flat space, chiral and anti-chiral multiplets can be obtained from a general multiplet~\genSfourd\ by imposing supersymmetric constraints:
\medskip
\item{$\bullet$} {\it Chiral Multiplet:} Imposing~$\t \chi^\alphadot = 0$ leads to a chiral multiplet~$\Phi=(\phi, \psi_\alpha, F)$ of~$R$-charge~$\fr$. Consistency with the transformation rules~\fdgmul\ shows that~$\Phi$ is embedded in a general multiplet~\genSfourd\ as follows,
\eqn\embedchiralfd{
\Phi = \left(\phi, -\sqrt2 i \psi_\alpha , 0, -2 i F, 0, - i D_\mu \phi, 0, 0, {r\over 4}(R-6 V_\mu V^\mu ) \phi \right)\, . 
}
The resulting supersymmetry transformations for the components of~$\Phi$ are given by
\eqn\susychiralfd{\eqalign{
&\delta \phi = \sqrt2 \zeta \psi~,\cr
& \delta \psi = \sqrt 2 \zeta F + \sqrt 2 i \sigma^\mu \t \zeta \, D_\mu \phi~, \cr
& \delta F =\sqrt 2 i D_\mu \big(\t \zeta \,\t \sigma^\mu \psi\big)~.}}

\medskip
\item{$\bullet$} {\it Anti-Chiral Multiplet:} The multiplet conjugate to~$\Phi$ is an anti-chiral multiplet~$\t \Phi = (\t \phi,\t \psi^{\dot\alpha}, F)$ of~$R$-charge~$-r$. It is embedded in a general multiplet~\genSfourd\ of~$R$-charge~$-r$ with~$\chi_\alpha = 0$,\eqn\embedantichiralfd{
\t\Phi =  \left(\t \phi, 0,\sqrt2 i \t\psi^\alphadot , 0,2 i\t F,  i D_\mu \t\phi, 0, 0, {r\over 4}(R-6 V_\mu V^\mu ) \t\phi \, \right)~.
}
Its supersymmetry transformations are given by
\eqn\susyantichiralfd{\eqalign{
&\delta \t\phi = \sqrt2 \t\zeta \t\psi~,\cr
& \delta \t\psi = \sqrt 2 \t\zeta \t F + \sqrt 2 i \t \sigma^\mu  \zeta \, D_\mu \t\phi~, \cr
& \delta \t F =\sqrt 2 i D_\mu \big( \zeta \sigma^\mu \t \psi\,\big)~.}}

\subsec{Gauge Fields and Charged Matter}

In order to describe gauge fields, we need to specify a compact gauge group~$G$ with Lie algebra~$\frak g$. Let~$T^a~(a = 1, \ldots, \dim G)$ denote a set of Hermitian generators for~$\frak g$ in the adjoint representation, normalized so that
\eqn\gennorm{
\Tr \left(T^a T^b\right) = \delta^{ab}~.
}
As in flat space, a vector multiplet~$\CV = \CV^a T^a$ is a general multiplet of vanishing~$R$-charge, which is valued in the adjoint representation of~$\frak g$ and subject to the gauge freedom
\eqn\defCVfd{
e^{-2 \CV'} = e^{i \t\Omega} e^{-2 \CV} e^{- i \Omega}~.
}
Here~$\Omega = \Omega^a T^a$ and~$\t \Omega = \t \Omega^a T^a$ are arbitrary adjoint-valued chiral and anti-chiral multiplets of vanishing~$R$-charge. The exponentials in~\defCVfd\ should be interpreted as infinite power series, where each term is evaluated according to the product rules in appendix~A.

We can use \defCVfd\ to fix Wess-Zumino (WZ) gauge, in which 
\eqn\VWZfd{
\CV= \left(0,0,0,0,0, a_\mu, \lambda_\alpha, \t \lambda^\alphadot, D\right)~.
}
All components of~$\CV$ are~$\frak g$-valued fields in the adjoint representation, e.g.~$a_\mu = a_\mu^a T^a$. The residual gauge freedom is parametrized by~$\Omega= \t\Omega= (\omega, 0, 0)$. For infinitesimal~$\omega$, 
\eqn\resgaugetransfofd{
\delta_\omega  a_\mu = \d_\mu \omega + i [\omega, a_\mu]~, \quad
\delta_\omega \lambda =  i [\omega,\lambda]~, \quad
\delta_\omega \t\lambda =  i [\omega,\t \lambda]~, \quad
\delta_\omega D =  i [\omega, D]~.
}
As expected, $a_\mu$ transforms like a~$\frak g$-valued gauge field, while the other components transform in the adjoint representation of~$\frak g$.  

Supersymmetry transformations do not preserve WZ-gauge, which must be restored by a compensating gauge transformation. This leads to the following transformation rules:
\eqn\gaugemultfd{\eqalign{
& \delta a_\mu = i (\zeta \sigma_\mu \t \lambda + \t \zeta\, \t \sigma_\mu \lambda)~,\cr
& \delta \lambda = i \zeta D +  \sigma^{\mu\nu}\zeta \, f_{\mu\nu}~, \cr
& \delta \t \lambda = - i \t \zeta D +  \t \sigma^{\mu\nu} \t \zeta \,  f_{\mu\nu} ~,\cr
& \delta D = - D_\mu (\zeta \sigma^\mu \t \lambda - \t \zeta \, \t \sigma^\mu \lambda) + 2 i V_\mu (\zeta \sigma^\mu \t \lambda + \t \zeta \, \t \sigma^\mu \lambda)~.}}
Here the adjoint-valued field strength~$f_{\mu\nu}$ is given by
\eqn\fmunufd{
f_{\mu\nu} =  \d_\mu a_\nu - \d_\nu a_\mu - i [a_\mu, a_\nu]~.
}
The covariant derivative~$D_\mu$ acts as in~\covddef, supplemented by the usual minimal terms to ensure gauge covariance under~\resgaugetransfofd. For instance,
\eqn\dexmple{
D_\mu \lambda = \grad_\mu\lambda - i \Big(A^{(R)}_\mu+{3\over 2} V_\mu\Big) \lambda - i[a_\mu,\lambda]~.}
The Lie derivative~$\CL'_K$ in~\modlie\ is similarly modified, which ensures gauge covariance of the supersymmetry algebra~\susyalg\ and consistency with the transformation rules~\gaugemultfd. 

As in flat space, we can consider a field-strength multiplet~$\Lambda_\alpha$, whose bottom component is the gaugino~$\lambda_\alpha$. We will only need its gauge-invariant square~$\Tr\,\Lambda^2$, which is a chiral multiplet whose components~$\left(\phi_{\Lambda^2}, \psi_{\Lambda^2}, F_{\Lambda^2}\right)$ are given by
\eqn\Wsqchiral{\eqalign{
&\phi_{\Lambda^2} = \Tr\,  \lambda \lambda~, \cr
& \psi_{\Lambda^2} = \sqrt 2 \Tr \left(i \lambda D - \sigma^{\mu\nu}\lambda f_{\mu\nu}\right)~, \cr
& F_{\Lambda^2} = \Tr\left(  \half f_{\mu\nu}f^{\mu\nu} + {1\over 4} \ep^{\mu\nu\rho\sigma} f_{\mu\nu}f_{\rho\sigma} - D^2 + 2 i \lambda \sigma^\mu \Big(D_\mu -{3i\over 2}V_\mu\Big)\t\lambda\right)~.
}}
The components of the conjugate anti-chiral multiplet~$\Tr \, \t \Lambda^2$  are given by
\eqn\Wsqantichiral{\eqalign{
&\t\phi_{\t \Lambda^2} = \Tr\,  \t\lambda \t\lambda~, \cr
&  \t\psi_{\t \Lambda^2} =\sqrt 2 \Tr\left(\t\lambda\t\sigma^{\mu\nu} f_{\mu\nu} - i \t\lambda D\right)~, \cr
&\t  F_{\t \Lambda^2} = \Tr\left( \half f_{\mu\nu}f^{\mu\nu} - {1\over 4} \ep^{\mu\nu\rho\sigma} f_{\mu\nu}f_{\rho\sigma} - D^2 - 2 i \Big(D_\mu +{3i\over 2}V_\mu\Big)\lambda \sigma^\mu\t\lambda\right) \cr
& \hskip20pt = F_{\Lambda^2} - \Tr \left(\half \ep^{\mu\nu\rho\sigma} f_{\mu\nu} f_{\rho\sigma} + 2 i \grad_\mu \big(\lambda \sigma^\mu \t \lambda\big)\right)~.
}}
Here the last line shows that~$\t F_{\t \Lambda^2}$ and~$F_{\Lambda^2}$ only differ by a topological term, as well as a genuine total derivative. Note that~\Wsqchiral\ and~\Wsqantichiral\ do not follow from the multiplication rules in appendix~A, which only apply to chiral and anti-chiral multiplets whose bottom components are scalars. 

In the presence of gauge fields, the transformation rules~\susychiralfd\ and~\susyantichiralfd\ for charged chiral and anti-chiral matter fields are modified. Consider a chiral multiplet~$\Phi$ of~$R$-charge~$r$ transforming in some representation~${\frak R}$ of the gauge group with Hermitian generators~$T^a_{\frak R}$. The conjugate anti-chiral multiplet~$\t \Phi$ of~$R$-charge~$-r$ transforms in the representation~$\b{\frak R}$. As usual, we can view~$\Phi$ as  a column vector and let gauge transformations~\defCVfd\ act from the left, while~$\t \Phi$ is a row vector with gauge transformations acting from the right,
\eqn\gaugetransfo{
\Phi' = e^{i \Omega} \Phi~, \qquad \t \Phi' = \t \Phi e^{- i \t \Omega}~,
}
where~$\Omega = \Omega^a T^a_{\frak R}$ and~$\t \Omega = \t \Omega^a T^a_{\frak R}$ are both valued in the~$\frak R$-representation of~$\frak g$. As before, \gaugetransfo\ should be evaluated using a series expansion for the exponentials and the product rules in appendix~A.\foot{Note that the multiplication rules for the product of two chiral multiplets or the product of two anti-chiral multiplets are the same as in flat space.} Under the residual gauge freedom~\resgaugetransfofd, $\delta_\omega \Phi = i \omega^a T^a_{\frak R} \Phi$ and~$\delta_\omega \t \Phi = - i \omega^ a  \t \Phi T^a_{\frak R}$, as expected. 

The supersymmetry transformations of charged chiral and anti-chiral multiplets in WZ-gauge are given by~\susychiralfd\ and~\susyantichiralfd, followed by a compensating gauge transformation to restore WZ-gauge. For the chiral multiplet~$\Phi$ the resulting transformation rules are given by
\eqn\susychiralfdgauged{\eqalign{
&\delta \phi = \sqrt2 \zeta \psi~,\cr
& \delta \psi = \sqrt 2 \zeta F + \sqrt 2 i \sigma^\mu \t \zeta~D_\mu \phi~, \cr
& \delta F =\sqrt 2 i D_\mu \big(\t \zeta\,\t \sigma^\mu \psi\big) - 2i \t\zeta\t\lambda \phi~.}}
As in~\dexmple, the derivative~$D_\mu$ is both~$R$- and~$G$-covariant,
\eqn\chicovd{
D_\mu \phi = \d_\mu \phi - i r \Big(A_\mu^{(R)} + {3 \over 2}V_\mu\Big) \phi - i a_\mu^a T^a_{\frak R} \phi~.
} 
Likewise, the action of~$\t \lambda$ on~$\phi$ in the last term of~$\delta F$ should be understood in the~$\frak R$-representation, i.e.~$\t \lambda \phi = \t \lambda^a T_{\frak R}^a \phi$. 

Similarly, the supersymmetry transformations for the conjugate anti-chiral multiplet~$\t \Phi$ are given by
\eqn\susyantichiralfdgauged{\eqalign{
&\delta \t\phi = \sqrt2 \t\zeta \t\psi~,\cr
& \delta \t\psi = \sqrt 2 \t\zeta \t F + \sqrt 2 i \t \sigma^\mu  \zeta~D_\mu \t\phi~, \cr
& \delta \t F =\sqrt 2 i D_\mu \big( \zeta \sigma^\mu \t \psi\big) + 2i \zeta \lambda\t\phi~.}}
Now~$D_\mu$ and the action of~$\lambda$ on~$\t \phi$ in~$\delta \t F$ should be understood in the~$\b {\frak R}$-representation,
\eqn\chitcovd{
D_\mu \t \phi = \d_\mu \t \phi + i r \Big(A_\mu^{(R)} + {3 \over 2}V_\mu\Big) \t \phi + i a_\mu^a \left(T^a_{\frak R}\right)^* \t \phi~,
} 
and~$\lambda \t \phi = - \lambda^a \left(T^a_{\frak R}\right)^* \t \phi$. 

It is straightforward to adapt the preceding discussion to the case where~$G$ is a global flavor symmetry and the vector multiplet describes non-dynamical background gauge fields. A given configuration for the bosonic fields~$a_\mu, D$ with~$\lambda = \t \lambda = 0$ is compatible with supersymmetry if the variations~$\delta\lambda$ and~$\delta \t \lambda$ in~\gaugemultfd\ vanish. Since~$a_\mu, D$ are not dynamical and we are in Euclidean signature, it is natural to allow complex-valued field configurations. (See the discussion at the end of section~2.1.) The supersymmetry transformations of chiral and anti-chiral fields that transform under the flavor symmetry group~$G$ are given by~\susychiralfdgauged\ and~\susyantichiralfdgauged\ with~$\lambda = \t \lambda = 0$.

\subsec{Supersymmetric Lagrangians}

We can use the multiplets described above to construct supersymmetric Lagrangians. For our purposes, it is sufficient to limit the discussion to the curved-space analogues of standard~$D$- and~$F$-term Lagrangians:
\medskip
\item{1.)} {\it D-terms:} Consider a general multiplet~\genSfourd\ of~$R$-charge~$r=0$, which may itself be the product of various other multiplets.  According to~\fdgmul, the~$D$-component of this multiplet does not transform into a total derivative, unlike in flat space. However, it is easy to check that the following Lagrangian does,
\eqn\Dtermfd{
\SL_D = -\half \left(D - 2 a_\mu V^\mu\right)~. 
}
This is the curved-space analogue of the usual flat-space~$D$-term.
\medskip
\item{2.)} {\it F-Terms:} The analogue of the flat-space superpotential is a chiral multiplet~$W$ of $R$-charge~$r=2$, which is generally a composite field. It follows from~\susychiralfd\ that its~$F$-term~$F_W$, whose~$R$-charge vanishes, transforms into a total derivative under supersymmetry, and hence it can serve as a supersymmetric Lagrangian,
\eqn\chflag{\SL_F = F_W~.}
We will refer to~$\SL_F$ as a chiral~$F$-term. Similarly, it follows from~\susyantichiralfd\ that the supersymmetry variation of~$\t F_{\t W}$, which resides in the conjugate anti-chiral multiplet~$\t W$ of~$R$-charge~$r=-2$, is a total derivative. We will refer to such a term in the supersymmetric Lagrangian as an anti-chiral~$F$-term,
\eqn\achflag{
\SL_{\t F} = \t F_{\t W}~.
}
\medskip 

\subsec{Examples}

Consider a single chiral multiplet~$\Phi$ of~$R$-charge~$r$. As in flat space, the canonical kinetic Lagrangian~$\SL_{\t \Phi \Phi}$ for~$\Phi$ is obtained by applying the~$D$-term formula~\Dtermfd\ to the neutral superfield~$\t \Phi \Phi$, whose components can be computed using the multiplication rules in appendix~A. Up to a total derivative,
\eqn\freechiral{\eqalign{
\SL_{\t \Phi \Phi} =~&  D^\mu \t \phi D_\mu \phi + i \t \psi \t \sigma^\mu D_\mu \psi - \t F F -{r \over 4} \left(R- 6 V^\mu V_\mu\right) \t \phi \phi  \cr
& - i V^\mu \left(\t \phi D_\mu \phi - \phi D_\mu \t \phi \right) + \half V_\mu \t \psi \t \sigma^\mu \psi~.
}}
Here the covariant derivative~$D_\mu$ acts as in~\covddef. This precisely agrees with the formula obtained in~\FestucciaWS\ from the rigid limit of the corresponding supergravity Lagrangian.\foot{The general K\"ahler sigma model described in~\FestucciaWS\ can be obtained by applying~\Dtermfd\ to the multiplet whose bottom component is the K\"ahler potential~$K$.} The matter Lagrangian~\freechiral\ is independent of~$V_\mu$ if we set the~$R$-charge of the chiral field~$\Phi$ to its superconformal value~$r={2 \over 3}$. This is due to the fact that the operator coupling to~$V_\mu$ is redundant when the flat-space field theory is superconformal.

We would like to generalize~\freechiral\ to the case where~$\Phi$ transforms in a representation~$\frak R$ of the gauge group~$G$, with~$\t \Phi$ transforming in the conjugate representation~$\b {\frak R}$. We must now use~\susychiralfdgauged\ and~\susyantichiralfdgauged\ to compute the components of~$\t \Phi \Phi$. Alternatively, we can apply the product rules in appendix~A to the multiplet~$\t \Phi e^{-2 \CV} \Phi$, which is invariant under the full gauge freedom in~\defCVfd\ and~\gaugetransfo. The resulting Lagrangian is given by
\eqn\LagPhifd{\eqalign{
\SL^G_{\t \Phi \Phi} =~& D^\mu \t \phi D_\mu \phi + i \t \psi \t \sigma^\mu D_\mu \psi - \t F F  -{r \over 4} \left(R- 6 V^\mu V_\mu\right) \t \phi \phi  \cr
& - i V^\mu \left(\t \phi D_\mu \phi - \phi D_\mu \t \phi \right) + \half V_\mu \t \psi \t \sigma^\mu \psi \cr
& + \t \phi D \phi + \sqrt 2 i \left(\t \phi \lambda \psi - \t \psi \t \lambda \phi\right)~. 
}}
Here the covariant derivatives are modified as in~\chicovd\ and~\chitcovd. 

In order to write down Yang-Mills kinetic terms for the gauge fields, we proceed as in flat space and use the~$F$ and~$\t F$-terms of the chiral and anti-chiral superfields~$\Tr\,\Lambda^2, \Tr\, \t \Lambda^2$ in~\Wsqchiral\ and~\Wsqantichiral,
\eqn\LagYMwithtau{
\SL_{\rm YM}= {\tau \over 16 \pi i} F_{\Lambda^2} - {\b\tau \over 16 \pi i} \t F_{\t \Lambda^2}~,}
where~$\tau$ the holomorphic gauge coupling,
\eqn\hgc{\tau={\theta\over 2\pi}+{4\pi i \over g^2}~.}
For future reference, we use the last equation in~\Wsqantichiral\ to rewrite~\LagYMwithtau\ as
\eqn\lagymfchi{
\SL_{\rm YM} = \left({\tau - \b \tau \over 16 \pi i }\right) F_{\Lambda^2} + {\b \tau \over 32 \pi i} \ep^{\mu\nu\rho\sigma} \Tr \left(f_{\mu\nu} f_{\rho\sigma}\right)~,
}
up to a total derivative. Therefore, the Yang-Mills Lagrangian is a chiral~$F$-term, up to a term which is topological and holomorphic in~$\b \tau$. In components,\eqn\LagYMfd{
\SL_{\rm YM}= {1\over g^2 } \Tr \left( {1\over 4} f^{\mu\nu} f_{\mu\nu} -\half D^2  + i \t\lambda \t\sigma^\mu \big(D_\mu + {3i \over 2}V_\mu\big) \lambda  \right) - {i \theta \over 64 \pi^2 }~\ep^{\mu\nu\rho\sigma} \Tr \left(f_{\mu\nu}f_{\rho\lambda}\right)~,
}
again up to a total derivative. If~$G$ contains several factors, each will have its own holomorphic gauge coupling. Note that the explicit appearence of~$V_\mu$ in~\LagYMfd\ is canceled by the~$V_\mu$-dependent piece of the covariant derivative~\dexmple, so that~$\SL_{\rm YM}$ is~$V_\mu$-independent. This is due to the superconformal invariance of the classical Yang-Mills Lagrangian. 

If~$G$ contains Abelian factors, additional supersymmetric terms become available. An important example is the Fayet-Iliopoulos (FI) term, which can be obtained by applying the~$D$-term formula~\Dtermfd\ to an Abelian vector multiplet~\VWZfd\ in Wess-Zumino gauge,
\eqn\filag{
\SL_{\rm FI} = \xi_{\rm FI} \left(D - 2 a_\mu V^\mu\right)~.
}
This Lagrangian is invariant under small gauge transformations of~$a_\mu$, since~$V^\mu$ is covariantly conserved, $\grad_\mu V^\mu =0$, but if the gauge group is compact, invariance of~$\SL_{\rm FI}$ under large gauge transformations may require~$\xi_{\rm FI}$ to be quantized (see for instance~\refs{\AharonyDHA}).
There are also terms that mix Abelian gauge fields with the~$R$-symmetry gauge field, e.g.~$\ep^{\mu\nu\rho\lambda} \d_\mu A^{(R)}_\nu f_{\rho\lambda}$. For simplicity, we will omit such terms, as well as FI-Terms~\filag, from our discussion, but it is straightforward to incorporate them into the framework developed below.

\newsec{Twisted Holomorphic Theories on Complex Manifolds}

In the previous section, we constructed supersymmetric Lagrangians on a four-manifold~$\CM$ that was assumed to admit one or several solutions to the Killing spinor equations~\speq. As we will review below, the existence of a Killing spinor requires~$\CM$ to be a complex manifold~\refs{\DumitrescuHA,\KlareGN}. In this section, we will describe supersymmetric theories on~$\CM$ in terms of twisted variables that are adapted to its complex structure. These variables will be convenient for studying the dependence of the partition function~$Z_{\CM}$ on coupling constants and the geometry of~$\CM$.

\subsec{Killing Spinors and Complex Manifolds}

From now on, we will assume that~$\CM$ admits a solution~$\zeta_\alpha$ of~$R$-charge~$+1$ to the first Killing spinor equation in~\speq, and hence a supercharge~$Q$ of~$R$-charge~$-1$. (The corresponding supersymmetry transformations on fields are denoted by~$\delta$.) As was shown in~\refs{\DumitrescuHA,\KlareGN}, this requires~$\CM$ to possess a Hermitian structure, i.e.~an integrable complex structure~${J^\mu}_\nu$ with compatible Hermitian metric~$g_{\mu\nu}$. The relation between~$\zeta_\alpha$, which is everywhere non-zero, and the complex structure is given by
\eqn\cctl{{J^\mu}_\nu=-{2i\over |\zeta|^2} \zeta^\dagger {\sigma^\mu}_\nu \zeta~.}
Using Fierz identities and the Killing spinor equation~\speq, it can be checked that the right-hand side of~\cctl\ is in fact an integrable complex structure. We will therefore work in local holomorphic coordinates~$z^i~(i=1,2)$ adapted to~${J^\mu}_\nu$, in which the only non-vanishing components of the complex structure and the Hermitian metric are 
\eqn\nonzerocomp{{J^i}_j = i {\delta^i}_j~, \qquad {J^{\b i}}_{\b j} = - i {\delta^{\b i}}_{\b j}~, \qquad g_{i\b j}~.}

It was argued in~\DumitrescuHA\ that the supercharge~$Q$ corresponding to~$\zeta_\alpha$ transforms like a scalar under holomorphic coordinate changes. A crucial role was played by the nowhere vanishing two-form~$P_{\mu\nu} = \zeta \sigma_{\mu\nu} \zeta$, which is a section of~$L^2 \otimes \CK$. Here~$L$ is the line bundle of local~$U(1)_R$ transformations and~$\CK$ is the canonical line bundle of complex~$(2,0)$-forms on~$\CM$. Since~$P_{\mu\nu}$ is everywhere non-zero, the line bundle~$L^2 \otimes \CK$ is trivial. We can therefore identify~$L = {\CK}^{-\half}$, up to a trivial line bundle. This defines a holomorphic twist, under which fields of~$R$-charge~$r$ become sections of~$\CK^{-{r \over 2}}$. (Below, we will discuss the conditions under which~$\CK^{-{r \over 2}}$ is well defined.) A special case of this twist was used in~\refs{\JohansenAW,\WittenEV,\vyasphd} to study certain~$R$-symmetric~$\CN=1$ theories on K\"ahler manifolds.   

In order to write explicit formulas, we define
\eqn\locdefs{p = P_{12}~, \qquad s = p g^{- {1 \over 4}}~, \qquad g = \det \left(g_{\mu\nu}\right)~.}
For future use, note that
\eqn\poverz{{|p|^2\over |\zeta|^4} = 4 \sqrt g~,}
which follows from~\locdefs\ and the fact that~$P_{12} = \zeta \sigma_{12} \zeta$. Under a holomorphic coordinate change~$z'^i = z'^i(z)$, 
\eqn\holcc{p'(z') = p(z) \det\left({\d z'^i \over \d z^j}\right)~, \qquad s'(z') = s(z) \left(\det\left({\d z'^i \over \d z^j}\right)\right)^\half \left(\det\left({\b {\d z'^i \over \d z^j}}\right)\right)^{-\half}~. }
Therefore,~$s$ transforms by a phase. After the twist, this phase is compensated by a~$U(1)_R$ transformation, so that~$s$ behaves as a scalar under holomorphic coordinate changes. In a suitable holomorphic frame, the Killing spinor~$\zeta_\alpha$ only depends on~$s$ and therefore also transforms as a scalar. The explicit form of~$\zeta_\alpha$ in such a frame can be found in~\DumitrescuHA, but it will not be needed here.

Given a choice of complex structure~${J^\mu}_\nu$ and Hermitian metric~$g_{\mu\nu}$ on~$\CM$, as well as a nowhere vanishing complex~$s$, the background fields~$A_\mu^{(R)}$ and are~$V_\mu$ almost completely fixed. The former is given by
\eqn\fdbg{\eqalign{& A^{( R )}_\mu = \hat A_\mu + A^{\rm flat}_\mu - {1 \over 4} \left(2{\delta_\mu}^\nu - i {J_\mu}^\nu\right) \grad_\rho {J^\rho}_\nu~,\cr
&\hat A_i={i\over 8} \d_i \log g~, \qquad \hat A_{\b i}= -{i\over 8}\partial_{\b i} \log g~,\cr
& A^{\rm flat}_\mu = -{i \over 2} \d_\mu \log s~.}}
while~$V^\mu$ is only determined up to a covariantly conserved, anti-holomorphic vector~$U^{\b i}$,
\eqn\fdbgii{V^\mu=\ha \grad^\nu{J_\nu}^\mu+U^\mu~,\qquad U^i=0~, \qquad \grad_\mu U^\mu=0~.}
Note that~$s$ is only unique up to multiplication by a well-defined, nowhere vanishing complex function~$s_0$. Multiplying~$s$ by~$s_0$ has the effect of shifting~$A_\mu^{(R)}$ by a gauge transformation.  If the homotopy class of the map~$s_0:\CM \rightarrow \C^*$ is non-trivial, it is a large gauge transformation. This is innocuous for fields whose~$R$-charge is properly quantized, which are described by well-defined integer powers of~$\CK$ upon identifying~$L = \CK^{-\half}$. As we will see below, some complex manifolds allow fields whose~$R$-charges are not properly quantized. After the twist, they are described by non-integer powers of~$\CK$, whose definition requires a choice of homotopy class for~$s_0$.\foot{Consider the following example~\LawsonYR. If the complex manifold~$\CM$ is spin, there exist square-roots of the canonical bundle~$\CK$. In order to define~$\sqrt \CK$, we must choose a spin structure, or equivalently a flat~$\Z_2$ connection. In our context, this amounts to specifying whether the winding numbers of~$s_0$ around one-cycles of~$\CM$ are even or odd.} 

\subsec {Twisted Variables}

Consider the general multiplet~$S$ introduced in~\genSfourd, whose bottom component~$C$ is a complex scalar of~$R$-charge~$r$. We can use the Killing spinor~$\zeta_\alpha$ to define a multiplet~$\CS$ of twisted variables:

\eqn\defforms{\eqalign{
& \CC=p^{-{r\over 2}} C~, \qquad \quad\CX= ip^{-{r\over 2}} \zeta \chi~,\cr
& \CX_{ij} = p^{-{r \over 2}} \zeta\sigma_{ij} \zeta \; {\zeta^\dagger \chi \over |\zeta|^2 }~,\qquad\quad \CM_{ij} = p^{-{r \over 2}} \zeta \sigma_{ij} \zeta \; M~, \cr
&  \t \CX_{\b i} = {i p^{-{r\over 2}} \over 2 |\zeta|^2} \zeta^\dagger \sigma_{\b i} \t \chi~,\qquad\quad  \CA_{\b i} = p^{-{r \over 2}} a_{\b i} + i \d_{\b i} \CC~,\cr
& \CA_i = p^{-{r \over 2}} a_i + i \grad_i^c \CC + r \Big(V_i + \half U_i\Big)\CC~, \qquad\quad \t \CL_i = i p^{-{r \over 2}} \zeta \sigma_i \t \lambda~,\cr
&  \t \CM_{\b i \b j} = p^{-{r \over 2}} {\zeta^\dagger \sigma_{\b i \b j} \zeta^\dagger \over |\zeta|^4} \; \t M~, \qquad\quad \CL_{\b i \b j} = 2 p^{-{r \over 2}} {\zeta^\dagger \sigma_{\b i \b j} \zeta^\dagger \over |\zeta|^4 } \; \zeta \lambda - 4 \left(\d_{\b i}  \t \CX_{\b j} - \d_{\b j}  \t \CX_{\b i}\right)~,\cr
& \CL = - i {p^{-{r\over 2}} \over |\zeta|^2} \; \zeta^\dagger \lambda~, \quad\qquad \CD = p^{-{r \over 2}} D + J^{\mu\nu} \grad_\mu^c \CA^{\phantom{c}}_\nu \cr
& \hskip115pt \quad+ r \Big(V_i + \half U_i\Big) \CA^i - {r \over 4} \left(R - 6 V^\mu V_\mu\right) \CC~.
}}
\medskip
\noindent They have the following properties:
\medskip
\item{1.)} Since~$\zeta_\alpha$ and~$p$ are nowhere vanishing, the transformation~\defforms\ from conventional to twisted variables is invertible. Note that the fields retain their statistics, e.g.~$\CC$ is a boson while~$\CX$ is a fermion, since~$\zeta_\alpha$ is a commuting spinor.
\medskip
\item{2.)} The twisted variables are sections of~$\CK^{-{r \over 2}} \otimes \Lambda^{n,m}$, for suitable~$n,m$, where~$\Lambda^{n,m}$ is the bundle of complex~$(n,m)$-forms adapted to the complex structure~\cctl. In particular, all twisted variables have vanishing~$R$-charge. 
\medskip
\item{3.)} The change of variables~\defforms\ is covariant under holomorphic coordinate changes. To make this manifest, we have expressed all derivatives in terms of the Chern connection~$\grad_\mu^c$, which is compatible with the metric and the complex structure (see appendix~B for a review). Note that~$\grad_{\b i}^c = \d_{\b i}$ when acting on sections of~$\CK$.
\medskip
\item{4.)} The main advantage of the twisted variables is that they transform in a particularly simple way under the supercharge~$\delta$ corresponding to the Killing spinor~$\zeta$. Using~\fdgmul\ with~$\t \zeta = 0$, we find that each line~$(\CY, \CZ)$ of~\defforms\ transforms as follows,
\eqn\lmult{\delta \CY = \CZ~, \qquad \delta \CZ = 0~,}
so that~$\delta^2 = 0$. Explicitly, the twisted variables~\defforms\ are paired into the following collection of~$(\CY, \CZ)$-multiplets with supersymmetry transformations as in~\lmult, 
\eqn\lmultlist{(\CC, \CX)~, \quad (\CX_{ij}, \CM_{ij})~, \quad (\t \CX_{\b i}, \CA_{\b i})~, \quad (\CA_i, \t \CL_i)~, \quad (\t \CM_{\b i \b j}, \CL_{\b i \b j})~, \quad (\CL, \CD)~.}
Note that the components of a given~$(\CY, \CZ)$-multiplet transform as sections of the same bundle, e.g.~$\CC, \CX$ are both scalars. This explicitly shows that the supercharge~$\delta$ is a scalar under holomorphic coordinate changes, as was emphasized in~\DumitrescuHA. The simple transformation rules in~\lmult\ should be compared with those in~\fdgmul, which are significantly more complicated.
\medskip
\item{5.)} The~$R$-charge~$r$ of the multiplet is restricted by the requirement that~$\CK^{-{r \over 2}}$ exists, which is generally only the case if~$r \in 2 \Z$. If~$\CK$ is topologically trivial, $c_1(\CK) = 0$, then~$\CK^{-{r \over 2}}$ exists for all~$r \in \R$. However, as was discussed at the end of section~3.1, the definition of~$\CK^{-{r \over 2}}$ generally depends on a choice of homotopy class for~$s_0$.
\medskip 

As in section~2, we can take two twisted multiplets~$\CS_1, \CS_2$, whose bottom components~$\CC_1, \CC_2$ transform as sections of~$\CK^{-{r_1\over 2}}, \CK^{-{r_2 \over 2}}$ and construct a product multiplet~$\CS$, whose bottom component~$\CC = \CC_1 \CC_2$ transforms as a section of~$\CK^{-\half (r_1 + r_2)}$. The other components of~$\CS$ are expressed in terms of the components of~$\CS_1, \CS_2$ using the supersymmetry transformations~\lmult\ and~\lmultlist. The resulting multiplication rules for twisted multiplets are summarized in appendix~A. 

Having obtained the twisted version of a general multiplet, we will now construct twisted analogues of vector, chiral, and anti-chiral multiplets. 

\subsec{Twisted Vector Multiplet}

The twisted version of a vector multiplet in WZ-gauge is obtained by applying the change of variables~\defforms\ to~\VWZfd. This leads to a twisted multiplet with zero~$R$-charge, whose only non-vanishing fields are given by
\eqn\twistedgauge{\CA_\mu = a_\mu~, \qquad \t \CL_i~, \qquad \CL~, \qquad \CL_{\b i \b j}~, \qquad \CD = D + \half J^{\mu \nu} f_{\mu\nu}~.}
Here~$f_{\mu\nu}$ is the field-strength defined in~\fmunufd. As in~\VWZfd, all variables in~\twistedgauge\ are valued in the Lie algebra~$\frak g$.

Using the supersymmetry transformations~\gaugemultfd, we find the following transformation rules for the twisted variables in~\twistedgauge,
\eqn\transgauge{\eqalign{&\delta a_i = \t \CL_{ i}~, \qquad \delta \t \CL_{i}=0~,\cr
&\delta \CL= \CD~, \qquad\delta \CD= 0~,\cr 
&\delta \CL_{\b i \b j} = 4 f_{\b i \b j}~,\qquad \delta a_{\b i}=0~.}}
Note that the pairs~$(a_i, \t \CL_i)$, $(\CL, \CD)$, $(\CL_{\b i \b j}, 4 f_{\b i \b j})$ are~$(\CX, \CY)$-multiplets and transform according to~\lmult. For later use, it is convenient to express the supersymmetry variation of~$f_{\mu\nu}$ in terms of the Chern connection (see appendix~B),
\eqn\fvar{\eqalign{& \delta f_{ij} = D_i^c \t \CL_j - D_j^c \t \CL_i + i {\left(dJ\right)^k}_{ij} \t \CL_k~,\cr
& \delta f_{i\b j} = - \delta f_{\b j i} = - D_{\b j}^c \t \CL_i~, \cr
& \delta f_{\b i \b j} = 0~.}}
Here~$D_\mu^c$ is the gauge-covariant version of the Chern connection,
\eqn\gcc{D_\mu^c \t \CL_i = \grad_\mu^c \t \CL_i- i [a_\mu, \t\CL_i]~.}

The transformations~\transgauge\ immediately imply that supersymmetric configurations for background gauge fields correspond to holomorphic vector bundles~\ClossetVRA. Setting all fermionic variables~$\t \CL_i, \CL, \CL_{\b i \b j}$ and their supersymmetry variations to zero, we find
\eqn\holvecbg{f_{\b i \b j} = 0~, \qquad \CD = 0~.}
The first equation defines a holomorphic vector bundle over the complex manifold~$\CM$. Below, we will mostly focus on Abelian background gauge fields, for which the condition~$f_{\b i \b j} = \d_{\b i} a_{\b j} - \d_{\b j} a_{\b i}  =0$ defines a holomorphic line bundle. (See~\ClossetVRA\ and references therein for additional details.)

\subsec{Twisted Chiral Multiplet}

The twisted version of a chiral multiplet~$\Phi = \left(\phi,\psi_\alpha , F\right )$ with~$R$-charge~$r$ is obtained by substituting~\embedchiralfd\ into the change of variables~\defforms. The only non-vanishing twisted fields are given by
\eqn\chimul{\eqalign{& \CC = p^{-{r \over 2}} \phi~, \qquad \CX = \sqrt 2 p^{-{r \over 2}} \zeta \psi~,\cr
& \CX_{ij} = - \sqrt2 i p^{-{r \over 2}} \zeta \sigma_{ij} \zeta {\zeta^\dagger \psi \over |\zeta|^2}~, \qquad \CM_{ij} = - 2 i p^{-{r \over 2}} \zeta \sigma_{ij} \zeta F~.}}
The pairs~$(\CC, \CX)$ and~$(\CX_{ij}, \CM_{ij}$) transform according to~\lmult,
\eqn\chiraltt{\eqalign{& \delta \CC = \CX~, \qquad \delta \CX = 0~,\cr
&  \delta \CX_{ij} = \CM_{ij}~, \qquad \delta \CM_{ij} = 0~.}}
These formulas remain valid if~$\Phi$ couples to dynamical or background gauge fields, as long as we work in WZ-gauge.

\subsec{Twisted Anti-Chiral Multiplet}

The twisted version of an anti-chiral multiplet~$\t \Phi = (\t \phi, \t \psi^\alphadot , \t F)$ with~$R$-charge~$-r$ similarly follows from substituting~\embedantichiralfd\ into~\defforms, with~$r \rightarrow -r$ in the latter formula.  Now the non-vanishing twisted fields are given by
\eqn\formsach{\eqalign{&\t \CC=p^{r\over 2}\t \phi~,\qquad  \t \CX_{\b i}= -{p^{r\over 2}\over \sqrt 2 |\zeta|^2} \zeta^\dagger \sigma_{\b i} \tilde \psi~,\qquad \t \CM_{\b i \b j}=2ip^{r\over 2} {\zeta^\dagger \sigma_{\b i \b j} \zeta^\dagger\over |\zeta|^4} \t F~,\cr
& \CA_i = 2 i \grad_i^c \t \CC - 2 r \big(V_i + \half U_i\big) \t \CC~, \qquad \CA_{\b i} = 2 i \d_{\b i } \t \CC~, \qquad \CL_{\b i \b j} = - 4 \left(\d_{\b i} \t \CX_{\b j} - \d_{\b j} \t \CX_{\b i}\right)~.}}
Here we have denoted the twisted version of~$\t \phi$ by~$\t \CC$, in order to distinguish it from the twisted version~$\CC$ of the chiral field~$\phi$. 

The supersymmetry transformations of the independent fields~$\t \CC, \t \CX_{\b i},\t \CM_{\b i \b j}$ are obtained by substituting~\formsach\ into~\lmultlist\ and applying~\lmult,
\eqn\achiralt{\delta \t \CC = 0~, \qquad \delta \t \CX_{\b i} = 2 i \d_{\b i} \t \CC~, \qquad \delta \t \CM_{\b i \b j} = - 4 \left(\d_{\b i} \t \CX_{\b j} - \d_{\b j} \t \CX_{\b i}\right)~.}
Note that these fields cannot be decomposed into~$(\CY, \CZ)$-multiplets transforming as in~\lmult. Rather, the supercharge is represented by the~$\b\d$-operator and the relation~$\delta^2 = 0$ is a consequence of the fact that~$\b \d^2 = 0$. If~$\t \Phi$ transforms in the representation~$\b {\frak R}$ under some dynamical or background gauge group, the formulas in~\achiralt\ are modified using~\susyantichiralfdgauged,
\eqn\achiraltgauge{\delta \t\CC=0~, \qquad  \delta \t \CX_{\b i}=2 i \hat \d_{\b i} \t\CC~, \qquad \delta \t \CM_{\b i \b j}=-4 (\hat \d_{\b i} \t \CX_{\b j}-\hat \d_{\b j} \t \CX_{\b i})+2 \CL_{\b i \b j}\t\CC~.}
Here~$\hat \d_{\b i}=\d_{\b i} - i a_{\b i}$ is the gauge-covariant~$\b \d$-operator and the fields~$a_{\b i}, \CL_{\b i \b j}$ belong to the twisted vector multiplet under which~$\t \Phi$ is charged. As in the discussion around~\chitcovd, they act on the component fields of~$\t \Phi$ in the~$\b {\frak R}$-representation, e.g.~$\hat \d_{\b i} \t\CC = \d_{\b i} \t \CC + i a_\mu^a \left(T^a_{\frak R}\right)^* \t \CC$ and $\CL_{\b i \b j} \t \CC = - \CL^a_{\b i \b j} \left(T^a_{\frak R}\right)^* \t \CC$.

\subsec{Twisted Lagrangians}

We will now express the supersymmetric~$D$- and~$F$-term Lagrangians constructed in section~2.3 in terms of twisted variables:
\medskip
{\item{1.) } {\it $D$-Terms:} The Lagrangian~$\SL_D$ in~\Dtermfd\ is constructed from a well-defined general multiplet of vanishing~$R$-charge. Using the transformation~\defforms\ to twisted variables (with~$r=0$), we can rewrite the~$D$-term Lagrangian as follows,
\eqn\dlagrew{\sqrt g \SL_{D} = - {\sqrt g \over 2} \left(D - 2 a_\mu V^\mu\right) = - {\sqrt g  \over 2} \big(\CD - 2 \CA_{\b i} U^{\b i} \big) +  ({\rm total\;derivative})~.}
Here we have used the explicit form for~$V^\mu$ in~\fdbgii, where~$U^{\b i}$ was defined. Since the multiplet was assumed to be well defined, we can drop the total derivative. It follows from~\lmult\ and~\lmultlist\ that~$\CD = \delta \CL$ and~$\CA_{\b i } =  \delta \t\CX_{\b i}$, so that~$D$-terms are~$Q$-exact,
\eqn\determqex{\sqrt g \SL_{D} = \delta \left( - {\sqrt g \over 2}  \left(\CL - 2 \t \CX_{\b i} U^{\b i}\right)\right)~.}
Since~$\delta^2 = 0$, it is also manifestly supersymmetric. The fact that~\determqex\ is~$Q$-exact implies that the partition function~$Z_{\CM}$ does not depend on~$D$-term parameters, i.e.~couplings that multiply supersymmetric terms of the form~\dlagrew. This does not imply that we can set all such parameters to zero (or other unphysical values) in the original Lagrangian, since the resulting path integral may fail to converge, see for instance~\WittenZE.
\medskip
{\item{2.)} {\it Chiral~$F$-Terms:} The Lagrangian~$\SL_{F}$ in~\chflag\ is constructed from a chiral multiplet~$W$ of~$R$-charge~$r=2$. Converting to the twisted variables defined in~\chimul, we can rewrite this term as follows,
\eqn\fterm{\SL_{F}~\sqrt gd^4x = F_W ~ \sqrt{g}d^4 x ={i\over 16} \sqrt{g}~\CM_{i j} dz^i\wedge dz^j\wedge d\b z^{\b 1}\wedge d\b z^{\b 2}~.}
Here~$\CM_{ij}$ is a~$(2,0)$-form with coefficients in~$\CK^{-1}$, i.e.~a scalar. According to~\chiraltt, $\CM_{ij} = \delta \CX_{ij}$, so that 
\eqn\chifqex{\SL_{F}~\sqrt gd^4x = {i\over 16} \delta\left(\sqrt{g}~\CX_{i j} dz^i\wedge dz^j\wedge d\b z^{\b 1}\wedge d\b z^{\b 2}\right)~,}
which is~$Q$-exact. Therefore,~$Z_\CM$ also does not depend on chiral~$F$-term parameters.
\medskip
{\item{3.) } {\it Anti-Chiral~$F$-Terms:} The Lagrangian~$\SL_{\t F}$ in~\achflag\ is based on an anti-chiral multiplet~$\t W$ of~$R$-charge~$r=-2$. Converting to the twisted variables defined in~\formsach\ and using~\poverz, we can rewrite~$\SL_{\t F}$ as follows,
\eqn\tfterm{\SL_{\t F}~\sqrt gd^4x = \t F_{\t W} ~ \sqrt{g}d^4 x = -{i \over 64}~ \t \CM_{\b i \b j} d \b z^{\b i } \wedge d \b z^{\b j} \wedge dz^1 \wedge dz^2~.} 
Note that the factor of~$\sqrt g$ has disappeared. This is due to the fact that~$\t \CM_{\b i \b j}$ is a~$(0,2)$-form with coefficients in~$\CK$, i.e.~a four-form. The supersymmetry transformations~\achiralt\ show that~$\t \CM_{\b i \b j}$ is not~$Q$-exact. However, we can use them to compute
\eqn\achisusy{\delta \SL_{\t F}~\sqrt gd^4x = {i \over 8} d \left(\t \CX_{\b i} d \b z^{\b i} \wedge dz^1 \wedge dz^2\right)~,}
which shows that~$\SL_{\t F}$ is supersymmetric. We conclude that~$Z_\CM$ may depend holomorphically on anti-chiral~$F$-term parameters. 
\bigskip

The simple supersymmetry transformation properties of the twisted variables allow us to identify many additional terms that are supersymmetric with respect to the supercharge~$Q$. A large class is furnished by~$Q$-exact terms, which are trivially supersymmetric because~$Q^2 = 0$. Such terms need not descend from a supergravity Lagrangian or reduce to an ordinary super-Poincar\`e-invariant Lagrangian in flat space. A non-trivial example is provided by a generalization of the two-form mass studied in~\WittenEV. This term requires two ingredients:
\medskip
\item{$\bullet$} A holomorphic~$(2,0)$-form~$\omega_{ij}$, which satisfies~$\b \d \omega = 0$. We will denote its complex conjugate by~$\b \omega_{\b i \b j}$. 
\medskip
\item{$\bullet$} A twisted chiral multiplet~\chimul\ with~$r = 0$, so that~$\CM_{ij}$ is a~$(2,0)$-form, and the~$(0,2)$-form~$\t \CM_{\b i \b j}$ in the conjugate twisted anti-chiral multiplet~\formsach. 
\medskip
\noindent We can then add the following term to the action,
\eqn\somegadef{S_\omega = \int \left(\b \omega \wedge \CM + \omega \wedge \t \CM\right)~.}
According to~\chiraltt, $\CM_{ij} = \delta \CX_{ij}$, so that the~$\b \omega$-term is~$Q$-exact, and hence supersymmetric. As in the discussion around~\achisusy, $\t \CM_{\b i \b j}$ is not~$Q$-exact, but it follows from~\achiralt\ that its supersymmetry variation is a total~$\b \d$-derivative. Integrating by parts and using~$\b \d \omega = 0$, we see that the~$\omega$-term is also supersymmetric. In the remainder of this paper, we will focus on Lagrangians consisting of conventional~$D$- and~$F$-terms.

\subsec{Examples}

Consider first the Lagrangian~$\SL_{\t \Phi \Phi}$ in~\freechiral\ for a chiral multiplet of~$R$-charge~$r$. In terms of the twisted chiral and anti-chiral multiplets in~\chimul\ and~\formsach, it is given by
\eqn\neuchirlag{\eqalign{\SL_{\t \Phi \Phi} =~& 2 g^{i \b j} \d_{\b j} \t \CC \Big(\grad_i^c - i r \big(V_i + \half U_i\big)\Big)\CC - {1 \over 8} \t \CM_{\b i \b j} \CM^{\b i \b j} + \d_{\b i } \t \CX_{\b j} \CX^{\b i \b j} \cr
& + i \t \CX^i \Big(\grad_i^c - i r \big(V_i + \half U_i\big)\Big) \CX + U^{\b i} \Big(2 i \d_{\b i} \t \CC \CC - \t \CX_{\b i} \CX\Big)~.
}}
Since this Lagrangian is a~$D$-term, it follows from~\determqex\ that it is~$Q$-exact. This can be made explicit by using the supersymmetry transformations in~\chiraltt\ and~\achiralt\ to rewrite~\neuchirlag\ as
\eqn\qexchilag{\SL_{\t \Phi \Phi} = \delta \left(- i \t \CX^i \Big(\grad_i^c - i r \big(V_i + \half U_i\big)\Big)\CC-{1 \over 8} \t \CM_{\b i \b j} \CX^{\b i  \b j} +  U^{\b i} \t \CX_{\b i} \CC\right)~.
}

Similarly, the Lagrangian~\LagPhifd\ for a chiral multiplet of~$R$-charge~$r$ transforming in the~$\frak R$-representation of the gauge group~$G$ can be expressed in terms of twisted chiral and anti-chiral multiplets, as well as the twisted vector multiplet~\twistedgauge,
\eqn\lagform{\eqalign{\SL^G_{\t \Phi \Phi} =~& 2 g^{i \b j} \hat \d_{\b j} \t \CC \Big(D^c_i - i r \big(V_i + \half U_i\big)\Big)\CC - {1 \over 8} \t \CM_{\b i \b j} \CM^{\b i \b j} + \hat \d_{\b i } \t \CX_{\b j} \CX^{\b i \b j} \cr
& + i \t \CX^i \Big(D_i^c - i r \big(V_i + \half U_i\big)\Big) \CX + U^{\b i} \Big(2 i \hat \d_{\b i} \t \CC \CC - \t \CX_{\b i} \CX\Big)\cr
& + \t \CC \CD \CC - \t \CC \CL \CX +{1 \over 4} \t \CC \CL_{\b i \b j} \CX^{\b i \b j} + \t \CX^i \t \CL_i \CC~.}}
Here~$D^c_i$ is the gauge-covariant version of the Chern connection~\gcc\ and~$\hat \d_{\b i}$ is the gauge-covariant~$\b \d$-operator introduced in~\achiraltgauge. For instance, 
\eqn\covcfmla{D^c_{i} \CC = \grad_i^c \CC - i a^a_i T^a_{\frak R} \CC~, \qquad \hat \d_{\b i} \t \CC = \d_{\b i} \t \CC + i a_{\b i}^a \left(T^a_{\frak R}\right)^* \t \CC~.}
The products in the last line of \lagform\ are interpreted as in~\LagPhifd. This Lagrangian is also a~$D$-term, which can be explicitly written in~$Q$-exact form using the supersymmetry transformations~\chiraltt\ and~\achiraltgauge,
\eqn\qexchilaggauge{\SL^G_{\t \Phi \Phi} = \delta \left(- i \t \CX^i \Big(D_i^c - i r \big(V_i + \half U_i\big)\Big)\CC -{1 \over 8} \t \CM_{\b i \b j} \CX^{\b i  \b j}+   U^{\b i} \t \CX_{\b i} \CC + \t \CC \CL \CC\right)~.
}

Finally, the Yang-Mills Lagrangian~\LagYMfd\ can be written in terms of the twisted vector multiplet~\twistedgauge,
\eqn\lagaform{\eqalign{\SL_{\rm YM} = {\tau-\b \tau\over 8\pi i }\Tr \Big( & f^{ij} f_{ij} - {1\over 2} \CD^2 + i \CD g^{i \b j} f_{i \b j} - \half \CL^{ij} D_i^c \t \CL_j - {i \over 4} \left(dJ\right)_{ij\b k} \CL^{ij} \t \CL^{\b k} \cr
& + i g^{i\b j} \CL D^c_{\b j} \t \CL_i \Big) + {\b \tau\over 32\pi i } \ep^{\mu\nu\rho\sigma} \Tr \left(f_{\mu\nu} f_{\rho\sigma}\right)~.}}
Here~$D_\mu^c$ acts as in~\gcc\ and~$\tau$ is the holomorphic gauge coupling~\hgc. It follows from~\lagymfchi\ that~$\SL_{\rm YM}$ is a chiral~$F$-term, and hence~$Q$-exact, up to a topological term that depends holomorphically on~$\b \tau$. We can see this explicitly by using~\transgauge\ and~\fvar\ to rewrite~\lagaform\ as follows:
\eqn\gauqexact{\eqalign{ \SL_{\rm YM}=\; & {\tau-\b \tau\over 8\pi i } \delta\left( \Tr \left({1 \over 4} \CL^{ij} f_{ij} - \half \CL \CD + i g^{i\b j} \CL f_{i \b j}\right)\right) \cr
& - \left(2 \pi i \b \tau\right) \left({1 \over 64\pi^2 } \ep^{\mu\nu\rho\lambda} \Tr \left(f_{\mu\nu} f_{\rho\lambda}\right)\right)~.}}
The last term in parentheses, which multiplies~$2 \pi i \b \tau$, is the Pontryagin density, which integrates to the instanton number. Hence, its contribution to the action does not depend on the complex structure or the Hermitian metric.

\newsec{Dependence of the Partition Function on the Geometry of $\CM$}

\subsec{Preliminaries}

In the previous section, we expressed supersymmetric theories on~$\CM$ in terms of twisted variables adapted to its complex geometry. This led to a simple proof that~$D$-terms and chiral~$F$-terms are~$Q$-exact, and hence the supersymmetric partition function~$Z_\CM$ cannot depend on coupling constants that multiply such terms. In this section we will examine the dependence of~$Z_\CM$ on the geometry of~$\CM$, i.e.~its complex structure and Hermitian metric, as well as holomorphic line bundles corresponding to Abelian background gauge fields. Using twisted variables, we will show that~$Z_\CM$ is independent of the Hermitian metric, and that it depends holomorphically on complex structure and line bundle moduli. This constitutes an alternative, more explicit,  proof of the results obtained in~\ClossetVRA.

As in~\ClossetVRA, we will consider infinitesimal deformations in the geometry of~$\CM$ and examine the associated change~$\Delta \SL$ of the Lagrangian. Deformations that lead to a~$Q$-exact~$\Delta \SL$ do not affect~$Z_\CM$. As was emphasized in the introduction, these arguments are essentially classical and may require modification (or break down) in the presence of quantum anomalies. Note that~$Q$-exact terms in the Lagrangian, such as~$D$-terms and chiral~$F$-terms, do not necessarily lead to~$Q$-exact terms in~$\Delta \SL$, since the supercharge is defined with respect to a fixed background and need not commute with all deformations. Therefore, $Q$-exact terms in the Lagrangian can contribute to the dependence of~$Z_{\CM}$ on the geometry of~$\CM$, even though changing their coefficients while keeping the geometry fixed has no effect. (Recall from the discussion after~\determqex\ that it is not legitimate to set these coefficients to zero, or other unphysical values.) 

In order to determine~$\Delta \SL$, we must decide how to vary the dynamical and background fields as we deform the geometry. This will be discussed in detail below. Here we will illustrate some of the possible subtleties in a simple example. Consider a field theory coupled to a Riemannian background metric~$g_{\mu\nu}$. In order to determine the energy-momentum tensor~$T_{\mu\nu}$, we can examine the change in the Lagrangian under a deformation~$\Delta g_{\mu\nu}$ of the background metric, $\Delta \SL = - \half \Delta g^{\mu\nu} T_{\mu\nu}$. We typically compute this variation while holding all fields other than the metric fixed, but this may not always be legitimate. For dynamical fields it is usually justified using their equations of motion, since any allowed field variation gives rise to terms that vanish on shell. This argument does not apply to background fields, but their effects can often be studied at fixed~$g_{\mu\nu}$. 

An exception arises if some dynamical or background fields obey constraints that depend on the metric. In order to preserve such constraints, we must also change the fields whenever we deform the metric. An instructive example arises in twisted~$\CN=2$ gauge theory on a four-manifold~\WittenZE, which contains a dynamical self-dual two-form field~$\chi_{\mu\nu} = \half \ep_{\mu\nu\rho\lambda} \chi^{\rho\lambda}$. Since the self-duality constraint depends on the metric, a deformation~$\Delta g_{\mu\nu}$ must be accompanied by a corresponding~$\Delta \chi_{\mu\nu}$, so that~$\chi_{\mu\nu} + \Delta \chi_{\mu\nu}$ is self-dual with respect to~$g_{\mu\nu} + \Delta g_{\mu\nu}$. This completely determines the anti-self-dual part of~$\Delta \chi_{\mu\nu}$, which is linear in~$\Delta g_{\mu\nu}$. The self-dual part of~$\Delta \chi_{\mu\nu}$ is arbitrary, but it gives rise to terms that vanish on shell, and hence it can be set to zero~\WittenZE. If we instead consider a situation where~$\chi_{\mu\nu}$ is a self-dual background field, the anti-self-dual part of~$\Delta \chi_{\mu\nu}$ is still fixed in terms of~$\Delta g_{\mu\nu}$, but the self-dual part of~$\Delta\chi_{\mu\nu}$ now constitutes a non-trivial deformation of the theory -- albeit one that can be studied while keeping the metric fixed. 

The procedure outlined above is model dependent: we need to know what fields appear in the Lagrangian~$\SL$ and how they change when we deform the geometry of~$\CM$. By contrast, the analysis in~\ClossetVRA\ did not require a Lagrangian description, since linearized deformations around flat space are parameterized by the supercurrent multiplet of the flat-space field theory in a model-independent way. Here we will focus on a large class of renormalizable theories with chiral and vector multiplets. In twisted variables, the kinetic terms for the chiral multiplets are given by~\neuchirlag\ or~\lagform, while the vector multiplets have Yang-Mills kinetic terms~\lagaform. We also include a superpotential, i.e.~a sum over products of elementary twisted chiral matter multiplets, which are multiplied using the rules in appendix~A, which gives rise to a chiral~$F$-term Lagrangian~\fterm. Similarly, the conjugate anti-chiral~$F$-term~\tfterm\ is constructed out of elementary twisted anti-chiral matter multiplets. For brevity, we will not discuss FI-terms~\filag\ or terms such as~\somegadef, although it is straightforward to incorporate them.

\subsec{Hermitian Metric}

Consider an infinitesimal deformation~$\Delta g_{i \b j}$ of the Hermitian metric, and the corresponding change~$\Delta \SL$ in the Lagrangian. Note the following:
\medskip
\item{1.)} The components of the twisted vector, chiral, and anti-chiral multiplets in~\twistedgauge, \chimul, and~\formsach\ do not satisfy any constraints that depend on the Hermitian metric. Hence, we can keep them fixed as we vary~$g_{i \b j}$. 
\medskip
\item{2.)} The only background field subject to a metric-dependent constraint is~$U^\mu$ in~\fdbg, which satisfies~$\grad_\mu U^\mu = 0$. We must therefore accompany~$\Delta g_{i\b j}$ by a change~$\Delta U^\mu$, whose precise form will not be important. 
\medskip
\item{3.)} The supersymmetry transformations~\transgauge, \chiraltt, and~\chiraltt\ or~\achiraltgauge\ for twisted vector, chiral, and anti-chiral multiplets do not depend on~$g_{i\b j}$ or~$U^\mu$. In particular, the derivatives that appear in the supersymmetry variations~\chiraltt\ or~\achiraltgauge\ for twisted anti-chiral multiplets only involve the (gauge-covariant)~$\b \d$-operator. \medskip
\noindent Therefore, supersymmetry transformations commute with the deformation~$\Delta g_{i \b j}$. This means that~$Q$-exact terms in the Lagrangian lead to~$Q$-exact terms in~$\Delta \SL$, which do not affect~$Z_\CM$. This applies to the~$D$-terms and chiral~$F$-terms in~\determqex\ and~\chifqex, as well as the Yang-Mills Lagrangian~\gauqexact, which is $Q$-exact up to a topological term.

We must now examine the anti-chiral~$F$-term Lagrangian~\tfterm, which is not~$Q$-exact. In our setup, the only such term is the anti-chiral part of the superpotential. It is constructed from elementary twisted anti-chiral superfields~\susyantichiralfd\ using the multiplication rules in appendix~A, which do not depend on the Hermitian metric~$g_{i \b j}$ or on~$U^\mu$. This implies that the twisted variable~$\t \CM_{\b i \b j}$ in~\tfterm\ does not change when we vary~$g_{i\b j}$, because the elementary twisted anti-chiral fields are held fixed. Since there is no other source of metric dependence in~\tfterm, this term does not change as we vary the Hermitian metric.\foot{Note that this argument does not apply to the anti-chiral~$F$-term part of the Yang-Mills Lagrangian~\lagaform, which is proportional to~$\b \tau$. This term is not a product of elementary twisted anti-chiral multiplets~\formsach\ and it explicitly depends on the Hermitian metric~$g_{i \b j}$. However, as we have explained above, it is~$Q$-exact up to a topological term, which does not depend on~$g_{i \b j}$.}

We conclude that the partition function~$Z_{\CM}$ does not depend on the Hermitian metric. Our proof generalizes the discussion in~\JohansenAW, which used twisted variables to argue for the metric independence of various observables in certain~$\CN=1$ theories on K\"ahler manifolds. Note that the need to vary~$U^\mu$ as we deform~$g_{i\b j}$, which was explained in point~$2.)$ above, did not play an essential role. This is due to the fact that~$U^\mu$ only appears inside~$Q$-exact terms and that the supersymmetry transformations do not depend on it. Therefore, the partition function~$Z_\CM$ does not depend on~$U^\mu$ either.\foot{This is consistent with the results of~\ClossetVRA, where it was shown that~$U^\mu$ can only affect the partition function~$Z_\CM$ if the flat-space theory does not possess a Ferrara-Zumino supercurrent multiplet. This does not happen for theories constructed out of conventional~$D$- and~$F$-terms, but it is known to occur in the presence of an FI-term~\filag.}

\subsec{Holomorphic Line Bundles}

As was explained around~\holvecbg, supersymmetric configurations for an Abelian background gauge field~$a_\mu$  satisfy~$f_{\b i \b j} = 0$ and hence define a holomorphic line bundle over the complex manifold~$\CM$. Consider an infinitesimal deformation~$a'_\mu = a_\mu + \Delta a_\mu$, where~$\Delta a_\mu$ is a well-defined one-form. In order for the~$(0,2)$-part of the new field strength to vanish, $\Delta a_{\b i}$ must satisfy 
\eqn\defsb{\d_{\b i} \big(\Delta a_{\b j}\big)-\d_{\b j} \left(\Delta a_{\b i}\right)=0~.}
Dividing by complexified background gauge transformations, we see that deformations of the holomorphic line bundle defined by~$a_\mu$ are parameterized by the Dolbeault cohomology class~$\left[\Delta a_{\b i}\right] \in H^{0,1}(\CM)$. Since~$\CM$ is compact, the dimension of~$H^{0,1}(\CM)$, which counts the number of bundle moduli, is finite.

Consider the change~$\Delta \SL$ in the Lagrangian induced by~$\Delta a_\mu$. As in the previous subsection, we keep all twisted variables, as well as all other background fields, fixed. Therefore, the anti-chiral~$F$-term superpotential~\tfterm\ does not change as we vary~$a_\mu$. We are left to examine the~$Q$-exact~$D$-terms and chiral~$F$-terms~\determqex\ and~\chifqex. While the supersymmetry transformations~\transgauge\ and~\chiraltt\ for dynamical vector and chiral multiplets are independent of~$a_\mu$, the transformation rules~\achiraltgauge\ for charged anti-chiral multiplets depend on~$a_{\b i}$, but not~$a_i$, through the gauge-covariant~$\b \d$-operator~$\hat \d_{\b i}$. Thus, supersymmetry transformations commute with the deformation~$\Delta a_\mu$, up to terms that only depend on~$\Delta a_{\b i}$. We conclude that~$Q$-exact terms in the Lagrangian lead to~$Q$-exact terms in~$\Delta \SL$, up to terms that depend on~$\Delta a_{\b i}$, but not on~$\Delta a_i$. This shows that~$Z_\CM$ is a locally holomorphic function of the line bundle moduli.\foot{As in~\ClossetVRA\ we are only considering infinitesimal deformations of the bundle moduli, which are not sensitive to the global structure of the moduli space, e.g.~its singularities.}

\subsec{Complex Structure}

An infinitesimal complex structure deformation takes the form~${{J'}^\mu}_\nu = {J^\mu}_\nu + \Delta {J^\mu}_\nu$. In order for~${J'^\mu}_\nu$ to be an integrable complex structure, $\Delta {J^\mu}_\nu$ must satisfy
\eqn\changeJ{\Delta {J^i}_j = \Delta {J^{\b i}}_{\b j} = 0~, \qquad \d_{\b j} \left( \Delta {J^{i}}_{\b k}\right)- \d_{\b k} \big( \Delta {J^i}_{\b j} \big) = 0~,}
where we continue to use holomorphic coordinates adapted to undeformed complex structure~${J^\mu}_\nu$. Infinitesimal diffeomorphisms induce deformations of the form~$\Delta {J^i}_{\b j} = \d_{\b j} \ep^i$, which do not change the complex structure. We must also deform the metric, $g'_{\mu\nu} = g_{\mu\nu} + \Delta g_{\mu\nu}$, so that it remains Hermitian with respect to the deformed complex structure,
\eqn\changemet{\Delta g_{ij} = {i \over 2} \left(g_{i \b k} \Delta {J^{\b k}}_j + g_{j \b k} \Delta {J^{\b k}}_i\right)~, \qquad \Delta g_{\b i \b j} = -{i \over 2} \left(g_{k \b i} \Delta {J^k}_{\b j} + g_{k \b j} \Delta {J^k}_{\b i}\right)~.}
Deformations of the form~$\Delta g_{i \b j}$, which are compatible with the undeformed complex structure, were discussed in section~4.2, and hence we can set them to zero here. 

As we saw in section~3, the definition of the supercharge~$Q$, the twisted variables, and their transformation rules crucially depends on the complex structure. We must carefully determine how they change under the deformation~$\Delta {J^\mu}_\nu$. This analysis, though greatly simplified by the use of twisted variables, is somewhat lengthy and will be presented in section~4.5. Here we will need the following two results:
\medskip
\item{1.)} When acting on twisted variables, the commutator~$[\Delta, \delta]$ of the complex structure deformation~$\Delta {J^\mu}_\nu$ and a supersymmetry transformation~$\delta$ does not depend on~$\Delta {J^{\b i}}_j$,
\eqn\comhol{
\left[\Delta, \delta\right] = {\rm holomorphic \hskip4pt in}\hskip4pt \Delta {J^i}_{\b j}~.
}

\item{2.)} Under the complex structure deformation~$\Delta {J^\mu}_\nu$, the following fields in a twisted anti-chiral multiplet~\formsach\ do not change,
\eqn\dletacmult{
\Delta \t \CC = \big(\Delta \t \CX\big)_{\b i} = \big(\Delta \CM\big)_{\b i \b j} = 0~.
}

\medskip

We will now prove that the partition function~$Z_{\CM}$ is a locally holomorphic function of the complex structure moduli. This follows immediately for~$Q$-exact terms in the Lagrangian, since we can use~\comhol\ to commute the complex structure deformation through the supercharge~$\delta$, at the expense of terms that are holomorphic in~$\Delta {J^i}_{\b j}$. Therefore, $Q$-exact terms in the Lagrangian give rise to~$Q$-exact terms in~$\Delta \SL$, up to terms holomorphic in~$\Delta {J^i}_{\b j}$. Again, this applies to~$D$-terms~\determqex\ and chiral~$F$-terms~\chifqex, as well as the Yang-Mills Lagrangian~\gauqexact, which is $Q$-exact up to a topological term that does not depend on the complex structure. 
 
As before, we must separately examine the anti-chiral~$F$-term superpotential~\tfterm, which is not~$Q$-exact. It is written in terms of a twisted variable~$\t \CM_{\b i \b j}$, which is constructed by combining elementary twisted anti-chiral multiplets according to the multiplication rules in appendix~A. These rules do not explicitly depend on the complex structure or any other background fields that may shift under~$\Delta {J^\mu}_\nu$. Since the components~\dletacmult\ of the elementary twisted anti-chiral fields are held fixed, the same is true for~$\t \CM_{\b i\b j}$. Therefore, the anti-chiral~$F$-term superpotential does not change as we deform the complex structure.

\subsec{Proof that~$\left[\Delta, \delta\right]$ is Holomorphic in~$\Delta {J^i}_{\b j}$}

We will now analyze how dynamical twisted vector, chiral, and anti-chiral multiplets, as well as their supersymmetry transformations, change when we deform the complex structure. Background fields require additional care and will be discussed at the end. This allows us to explicitly compute the commutator~$[\Delta, \delta]$ of the complex structure deformation~$\Delta {J^\mu}_\nu$ and a supersymmetry transformation~$\delta$, and hence to prove that it is holomorphic in~$\Delta {J^i}_{\b j}$ as stated in~\comhol. Along the way, we will also establish~\dletacmult.
\bigskip
\noindent{\it  Twisted Vector Multiplet: } 

The components of this multiplet and its supersymmetry transformations are given by~\twistedgauge\ and~\transgauge, respectively. The scalars~$\CL$,~$\CD$ and the one-form~$a_\mu$ do not satisfy any constraints, and hence we can hold them fixed as we deform the complex structure. However, $\t \CL_i$ is a~$(1,0)$-form and~$\CL_{\b i \b j}$ is a~$(0,2)$-form,
\eqn\jveccons{\eqalign{& \left({\delta^\nu}_\mu + i{J^\nu}_\mu\right) \t \CL_\nu = 0~, \cr
& \CL_{\mu\nu} = \CL_{[\mu\nu]}~, \qquad \left({\delta^\rho}_\mu - i{J^\rho}_\mu\right) \CL_{\rho\nu} = 0~.}}
As discussed in section~4.1, the fact that these constraints depend on~${J^\mu}_\nu$ requires us to accompany the deformation~$\Delta {J^\mu}_\nu$ by changes~$\t \CL'_\mu = \t \CL_\mu + \Delta \t \CL_\mu$ and~$\CL'_{\mu\nu} = \CL_{\mu\nu} +\Delta \CL_{\mu\nu}$. We can set~$\Delta \t \CL_{i} = \Delta \CL_{\b i \b j} = 0$, since they satisfy the constraints in~\jveccons\ and thus give rise to terms that vanish on shell. Solving for the remaining components of~$\Delta \t \CL_\mu$ and~$\Delta \CL_{\mu\nu}$, we find
\eqn\vecchanges{\eqalign{& (\Delta \t \CL)_{\b i} = -{i \over 2} \Delta {J^j}_{\b i} \t \CL_j~, \cr
& (\Delta \CL)_{i \b j} = - (\Delta \CL)_{\b j i} = {i \over 2} \Delta {J^{\b k}}_i \CL_{\b k \b j}~, \qquad (\Delta \CL)_{i j} = 0~.}}

In order to compute the commutator~$[\Delta, \delta ]$, it is helpful to express the transformation rules~\transgauge\ in a general coordinate system,
\eqn\gencoordtrans{\eqalign{& \delta a_\mu = \t \CL_\mu~, \qquad \delta \t \CL_\mu = 0~,\cr
& \delta \CL = \CD~, \qquad \delta \CD = 0~,\cr
& \delta \CL_{\mu\nu} = \left({\delta^\alpha}_\mu + i {J^\alpha}_\mu\right)\left({\delta^\beta}_\nu + i {J^\beta}_\nu\right) f_{\alpha\beta}~.}}
Keeping~$a_\mu$ fixed, i.e.~$\Delta a_\mu = 0$, this leads to
\eqn\acomm{[\Delta, \delta] a_\mu = \Delta(\delta a_\mu) = \Delta \t \CL_\mu~.}
Substituting~\vecchanges, we obtain
\eqn\acommcomp{\left([\Delta, \delta]a\right)_i = 0~, \qquad \left([\Delta, \delta] a\right)_{\b i} = - {i \over 2} {\Delta J^j}_{\b i} \t \CL_{j}~.}
Similarly, $\CL$ and~$\CD$ are fixed, $\Delta \CL = \Delta \CD = 0$, so that~\gencoordtrans\ implies
\eqn\ldcom{[\Delta, \delta] \CL = [\Delta, \delta] \CD = 0~.}
Finally, we use~\gencoordtrans\ to compute
\eqn\lformcom{[\Delta, \delta] \CL_{\mu\nu} = i \left(\Delta {J^\alpha}_\mu \left({\delta^\beta}_\nu + i {J^\beta}_\nu\right)  +  \left({\delta^\alpha}_\beta + i {J^\alpha}_\beta\right) \Delta {J^\beta}_\nu\right) f_{\alpha\beta} - \delta (\Delta \CL_{\mu\nu})~.}
Substituting for~$\Delta \CL_{\mu\nu}$ from~\vecchanges, we find
\eqn\lformcomcomp{\eqalign{& ([\Delta, \delta] \CL)_{\b i \b j} = 2 i \Delta {J^k}_{\b i} f_{k \b j} + 2 i \Delta {J^k}_{\b j} f_{\b i k}~, \cr
&  ([\Delta, \delta] \CL)_{i \b j} = ([\Delta, \delta] \CL)_{\b j i} =  ([\Delta, \delta] \CL)_{ij} = 0~.}}
The commutators in~\acommcomp, \ldcom, and~\lformcomcomp\ only depend on~$\Delta {J^i}_{\b j}$, but not its complex conjugate~$\Delta {J^{\b i}}_j$. This proves~\comhol\ when acting on twisted vector multiplets.

\bigskip

\noindent {\it Twisted Chiral Multiplet:} 

The supersymmetry transformations for a twisted chiral multiplet~\chimul\ of~$R$-charge~$r$ are given by~\chiraltt. The fields~$\CC$ and~$\CX$ are sections of~$\CK^{-{r\over 2}}$, where~$\CK$ is the canonical bundle corresponding to the complex structure~${J^\mu}_\nu$. Nevertheless, we will now argue that it is legitimate to hold~$\CC$ and~$\CX$ fixed as we deform~${J^\mu}_\nu$. For simplicity, we first consider the case~$r = -2$, so that~$\CC$ is a section of~$\CK$. It follows from the definition~\chimul\ of~$\CC$ in terms of the trivializing section~$p = P_{12}$ in~\locdefs\ that~$\CC = \CC_{12}$, where~$\CC_{ij}$ is a~$(2,0)$-form. Explicitly, $\CC_{\mu\nu} = P_{\mu\nu} \phi$, with~$\phi$ the scalar component of the original untwisted chiral multiplet, and hence~$\CC_{\mu\nu}$ satisfies 
\eqn\cantfcons{\CC_{\mu\nu} = \CC_{[\mu\nu]}~, \qquad \left({\delta^\rho}_\mu + i {J^\rho}_\mu\right) \CC_{\rho\nu} = 0~.}
In a coordinate patch, we can express the relationship between~$\CC_{\mu\nu}$ and~$\CC$ through the symbol~$\ep_{\mu\nu}$, which satisfies the constraints~\cantfcons\ and is normalized as~$\ep_{12} = 1$,
\eqn\cep{\CC_{\mu\nu} = \CC \ep_{\mu\nu}~.}
Thus~$\ep_{\mu\nu}$ is a local basis for sections of the line bundle~$\CK$, and~$\CC$ is the projection of~$\CC_{\mu\nu}$ onto this basis.  Due to the constraint in~\cantfcons, we must accompany the deformation~$\Delta {J^\mu}_\nu$ by a change~$\CC'_{\mu\nu} = \CC_{\mu\nu} + \Delta \CC_{\mu\nu}$, where
\eqn\deltacc{\left(\Delta \CC\right)_{ij} = 0~, \qquad \left(\Delta \CC\right)_{i \b j} = - \left(\Delta \CC\right)_{\b j i} = -{i \over 2} \Delta {J^k}_{\b j} \CC_{ik}~, \qquad \left(\Delta \CC\right)_{\b i \b j} = 0~.}
Since~$\ep_{\mu\nu}$ satisfies the same constraints as~$\CC_{\mu\nu}$, it must change in exactly the same way. It then follows from~\cep\ that the coefficient function~$\CC$ is held fixed as we deform the complex structure, i.e.~$\Delta \CC = 0$. A completely analogous discussion applies for arbitrary powers of~$\CK$, so that we can keep~$\CC$ and~$\CX$ fixed under the deformation~$\Delta {J^\mu}_\nu$ for any value~$r$ of the~$R$-charge,
\eqn\nochange{\Delta \CC = \Delta \CX =0~.}

The other two fields~$\CX_{ij}$, $\CM_{ij}$ in~\chimul\ are~$(2,0)$-forms with coefficients in~$\CK^{-{r\over 2}}$. Following the same logic, we treat these fields as~$(2,0)$-forms, i.e.~exactly as in~\deltacc, 
\eqn\deltaxm{\eqalign{& \left(\Delta \CX\right)_{ij} = 0~, \quad \left(\Delta \CX\right)_{i \b j} = - \left(\Delta \CX\right)_{\b j i} = -{i \over 2} \Delta {J^k}_{\b j} \CX_{ik}~, \quad \left(\Delta \CX\right)_{\b i \b j} = 0~,\cr
& \left(\Delta \CM\right)_{ij} = 0~, \quad \left(\Delta \CM\right)_{i \b j} = - \left(\Delta \CM\right)_{\b j i} = -{i \over 2} \Delta {J^k}_{\b j} \CM_{ik}~, \quad \left(\Delta \CM\right)_{\b i \b j} = 0~.}}
We can now use~\chiraltt, \nochange, and~\deltaxm\ to compute the commutator~$[\Delta, \delta]$ on the component fields of a twisted chiral multiplet,
\eqn\ddcomtc{[\Delta, \delta] \CC = [\Delta, \delta] \CX = \left([\Delta, \delta] \CX\right)_{\mu\nu} = \left([\Delta, \delta] \CM\right)_{\mu\nu} = 0~,}
so that~\comhol\ is trivially satisfied in this case.

\bigskip

\noindent{\it Twisted Anti-Chiral Multiplet:} 

The independent fields in a twisted anti-chiral multiplet~\formsach\ of~$R$-charge~$-r$ are~$\t \CC$, which is a section of~$\CK^{r \over 2}$,  a~$(0,1)$-form~$\t \CX_{\b i}$ with coefficients in~$\CK^{r \over 2}$, and a~$(0,2)$-form~$\t \CM_{\b i \b j}$ with coefficients in~$\CK^{r\over 2}$. As before, we can hold~$\t \CC$ fixed as we deform the complex structure, while~$\t \CX_\mu$ and~$\t \CM_{\mu\nu}$ change like~$(0,1)$-forms and~$(0,2)$-forms, 
\eqn\accc{\eqalign{& \Delta \t \CC = 0~, \cr
& \big(\Delta \t \CX\big)_{\b i} = 0~, \qquad \big(\Delta \t \CX\big)_i = {i \over 2} \Delta {J^{\b j}}_i \t \CX_{\b j}~,\cr
& \big(\Delta \t \CM\big)_{\b i \b j} = 0~, \qquad \big( \Delta \t \CM\big)_{i \b j} = - \big(\Delta \t \CM\big)_{\b j i} = {i \over 2} \Delta {J^{\b k}}_i \t \CM_{\b k \b j}~, \qquad \big(\Delta \t \CM\big)_{ij} = 0~,}}
so that we have established~\dletacmult. As in~\gencoordtrans, it is helpful to rewrite the supersymmetry transformations~\achiralt\ in an arbitrary coordinate system before computing the commutator~$[\Delta, \delta]$. Since these transformations contain~$\b \d$-operators acting on sections of the canonical bundle, their covariant version involves the Chern connection, which leads to a cumbersome calculation. Instead, we will work out the case~$r = 0$, where~$\t \CC$ is a scalar, $\t \CX_{\b i}$ a~$(0,1)$-form, and~$\t \CM_{\b i \b j}$ a~$(0,2)$-form, so that we can use conventional exterior derivatives. The results for general~$r$ can then be inferred using covariance. 

When~$r=0$, it is straightforward to express the supersymmetry transformations~\achiralt\ in general coordinates,
\eqn\scalatgencor{\eqalign{& \delta \t \CC = 0~,\cr
& \delta \t \CX_\mu = i \left({\delta^\nu}_\mu + i {J^\nu}_\mu\right) \d_\nu \t \CC~, \cr
& \delta \t \CM_{\mu\nu} = -\left({\delta^\alpha}_\mu + i {J^\alpha}_\mu\right) \left({\delta^\beta}_\nu + i {J^\beta}_\nu\right) \left(\d_\alpha\t \CX_\beta - \d_\beta \t \CX_\alpha\right)~.}}
Together with~\accc, this allows us to compute~$[\Delta, \delta]$ for the case~$r=0$,
\eqn\rzeroaccomm{\eqalign{& [\Delta, \delta] \t \CC = 0~, \cr
& ([\Delta, \delta] \t \CX)_i = 0~, \qquad ([\Delta, \delta]\t \CX)_{\b i} = - \Delta {J^j}_{\b i} \d_j \t \CC~,\cr
& ([\Delta, \delta] \t \CM)_{ij} = ([\Delta, \delta] \t \CM)_{i \b j} = ([\Delta, \delta] \t \CM)_{\b j i} = 0~, \cr
& ([\Delta, \delta] \t \CM)_{\b i \b j} = - 2 i \Delta {J^k}_{\b i} \d_k \t \CX_{\b j} + 2 i \Delta {J^k}_{\b j} \d_k \t \CX_{\b i}~.}}
For~$r= 0$, these formulas are covariant under holomorphic coordinate changes. In general~$\CC$, $\t \CX_{\b i}$, $\t \CM_{\b i \b j}$ are sections of~$\CK^{r \over 2}$, so that the derivatives in~\rzeroaccomm\ are no longer covariant.  This forces us to add additional terms, which are uniquely determined by dimensional analysis and covariance. The formulas in~\rzeroaccomm\ that must be modified  take the following form:
\eqn\accommcov{\eqalign{& ([\Delta, \delta]\t \CX)_{\b i} = - \Delta {J^j}_{\b i} { \d}_j \t \CC - {r \over 2} \d_j (\Delta {J^j}_{\b i}) \t \CC~,\cr
& ([\Delta, \delta] \t \CM)_{\b i \b j} = - 2 i \Delta {J^k}_{\b i}  \d_k \t \CX_{\b j} + 2 i \Delta {J^k}_{\b j}  \d_k \t \CX_{\b i} - i r \d_k (\Delta {J^k}_{\b i}) \t \CX_{\b j} + i r \d_k (\Delta {J^k}_{\b j}) \t \CX_{\b i}~.}}
It can be checked that these expressions transform covariantly under holomorphic coordinate changes. As before, we find that all commutators~$[\Delta, \delta]$ are holomorphic in~$\Delta {J^i}_{\b j}$. 

So far we have only discussed neutral twisted chiral and anti-chiral multiplets. The inclusion of dynamical gauge fields~$a_\mu$ is achieved by replacing~$\d_i \rightarrow \hat \d_i= \d_i-i a_i$ in~\accommcov, while the vanishing commutators in~\ddcomtc\ and~\rzeroaccomm\ are not modified. Background gauge fields that couple to global symmetries will be discussed below.

\bigskip

\noindent{\it Background Fields:} 

As explained in section~4.1, background fields that satisfy~${J^\mu}_\nu$-dependent constraints require additional care. For instance, the vector field~$U^\mu$ in~\fdbgii\ is anti-holomorphic with respect to~${J^\mu}_\nu$, i.e.~$U^i = 0$. We must therefore accompany the complex structure deformation~$\Delta {J^\mu}_\nu$ by a change in~$U^\mu$, but since~$U^\mu$ never enters the supersymmetry transformations, this does not affect the commutators~$[\Delta, \delta]$. 

A more interesting example is furnished by an Abelian background vector multiplet. Supersymmetry requires that the bosonic fields satisfy~\holvecbg,
\eqn\holvecebgrep{f_{\b i \b j} = 0~, \qquad \CD = 0~.}
The first equation defines a holomorphic vector bundle with respect to the complex structure~${J^\mu}_\nu$. We must therefore accompany~$\Delta {J^\mu}_\nu$ by a change~$\Delta a_\mu$ to ensure that~$a_\mu + \Delta a_\mu$ defines a holomorphic line bundle with respect to the new complex structure, while~$\CD = 0$ does not change. In order to determine~$\Delta a_\mu$ it is convenient to rewrite the first equation in~\holvecebgrep\ in a general coordinate system,
\eqn\genvb{\left({\delta^\alpha}_\mu + i{J^\alpha}_\mu\right)\left({\delta^\beta}_\nu + i {J^\beta}_\nu\right) \left(\d_\alpha a_\beta - \d_\beta a_\alpha\right) = 0~.}
Varying the complex structure, we find that~$\Delta a_i$ is unconstrained. Such deformations were shown to be~$Q$-exact in section~4.3, and hence we do not need to discuss them here. By contrast, $\Delta a_{\b i}$ satisfies the following constraint:
\eqn\daibcons{\d_{\b i} \big(\Delta a_{\b j}\big) - \d_{\b j} \left(\Delta a_{\b i}\right) = {i \over 2} \Delta {J^k}_{\b i} f_{\b j k} - {i \over 2} \Delta {J^k}_{\b j} f_{\b i k}~.}
Since~$\Delta a_{\b i}$ is a well-defined~$(0,1)$-form, this equation may not admit a solution if the~$(0,2)$-form on the right-hand side defines a non-trivial cohomology class in~$H^{0,2}(\CM)$. Such an obstruction typically signals the presence of a singularity in the moduli space of holomorphic line bundles, viewed as a fibration over the complex structure moduli space of the base manifold (see for instance~\Kobayashi). Here we will only discuss suitably generic points, where this obstruction is absent, so that~\daibcons\ determines~$\Delta a_{\b i}$ in terms of the complex structure deformation and the curvature of the undeformed background gauge field. Note that the right-hand side of~\daibcons\ is holomorphic in~$\Delta {J^i}_{\b j}$, and hence the same is true for~$\Delta a_{\b i}$. 

The fact that~$\Delta a_{\b i} \neq 0$ does not affect twisted vector and chiral multiplets, since~$a_\mu$ does not enter their supersymmetry transformations. However, it leads to a modification of the formulas in~\accommcov\ for twisted anti-chiral multiplets, 
\eqn\accgauge{\eqalign{& ([\Delta, \delta]\t \CX)_{\b i} = - \Delta {J^j}_{\b i} \left(\d_j - i a_j\right)\t \CC - {r \over 2} \d_j (\Delta {J^j}_{\b i}) \t \CC + 2 \Delta a_{\b i} \t \CC~,\cr
& ([\Delta, \delta] \t \CM)_{\b i \b j} = - 2 i \Delta {J^k}_{\b i} \left(\d_k - i a_k\right) \t \CX_{\b j} + 2 i \Delta {J^k}_{\b j} \left(\d_k - i a_k\right) \t \CX_{\b i} \cr
& \hskip62pt - i r \d_k (\Delta {J^k}_{\b i}) \t \CX_{\b j} + i r \d_k (\Delta {J^k}_{\b j}) \t \CX_{\b i} \cr
& \hskip62pt + 4 i \big(\Delta a_{\b i} \t \CX_{\b j} - \Delta a_{\b j} \t \CX_{\b i}\big)~,}}
since~\accommcov\ was derived for dynamical gauge fields, which satisfy~$\Delta a_\mu=0$. Unsurprisingly, the derivatives in~\accgauge\ have been rendered background gauge-covariant, but there are also additional terms that explicitly depend on~$\Delta a_{\b i}$. They do not spoil the fact that the commutators~$[\Delta, \delta]$ are holomorphic in~$\Delta {J^i}_{\b j}$, since~$\Delta a_{\b i}$ also has this property. This completes the proof of~\comhol.

\newsec{Dependence of the Partition Function on the~$R$-Symmetry}

In this section we will analyze the dependence of the partition function~$Z_\CM$ on the choice of~$R$-symmetry that enters the supersymmetric Lagrangian on~$\CM$. Here we phrase much of the discussion in the linearized language of~\ClossetVRA, but we will also show how to derive some key results using twisted variables. As an example, we discuss the~$R$-symmetry dependence of partition functions on primary Hopf surfaces, which are closely related to supersymmetric indices on~$S^3 \times S^1$. 

\subsec{Varying the~$R$-Symmetry}

In flat space, there is often more than one choice of~$U(1)_R$ symmetry. Given an Abelian flavor current~$j_\mu$, we can shift the~$R$-current~$j_\mu^{(R)}$ as follows,
\eqn\jmix{
j_\mu^{(R)} \rightarrow j_\mu^{(R)} + t  j_\mu~, \qquad t \in \R~.
}
Here we require both~$j_\mu^{(R)}$ and~$j_\mu$ to be conserved. The real parameter~$t$ quantifies the mixing between the two symmetries. Note the following:
\medskip
\item{1.)} There is an independent mixing parameter for every Abelian flavor symmetry, but in order to simplify the discussion, we will focus on a single~$t$.
\medskip
\item{2.)} In flat space, $t$ is arbitrary, but it may be restricted once we place the theory on a non-trivial manifold (see below). 
\medskip
\item{3.)} If the theory is superconformal, there is a unique choice of~$t$ that corresponds to the~$R$-symmetry residing in the superconformal algebra. It can be determined in flat space using~$a$-maximization~\IntriligatorJJ. 
\bigskip

We will now briefly review how~$j_\mu$,~$j_\mu^{(R)}$, and their superpartners couple to background fields at the linearized level around flat space, following the discussion and conventions of~\ClossetVRA. The flavor current~$j_\mu$ resides in a real linear multiplet, together with a scalar~$J$ and the fermions~$j_\alpha, \t j_{\alphadot}$. These operators couple to a background gauge field~$a_\mu$, a scalar~$D$ and the gauginos~$\lambda_\alpha, \t \lambda_\alphadot$,
\eqn\flcurbf{\SL_{\rm flavor} = a^\mu j_\mu + DJ + \lambda^\alpha j_\alpha + \t \lambda_\alphadot \t j^\alphadot + \cdots~,}
where the ellipsis denotes higher-order terms in the background fields, which are required by gauge invariance.\foot{A standard example is the seagull term~$a^\mu a_\mu \t \phi \phi$, which arises from the gauge-covariant derivatives in a scalar kinetic term~$D^\mu \t \phi D_\mu \phi$.} Supersymmetric configurations are determined by setting the gauginos~$\lambda_\alpha, \t \lambda_\alphadot$ and their supersymmetry variations to zero, which leads to~\holvecbg\ and the statement that~$a_\mu$ defines a holomorphic line bundle~$\CL$ over the complex manifold~$\CM$. The topology of~$\CL$ is fixed by its Chern class~$c_1(\CL) \in H^2(\CM, \Z)$, while its holomorphic structure is specified by a finite number of complex moduli. Locally, the moduli space is described by the Dolbeault cohomlogy~$H^{0,1}(\CM)$, but the moduli are typically subject to global identifications. 

The~$R$-current~$j_\mu^{(R)}$ resides in the same supermultiplet as the energy-momentum tensor~$T_{\mu\nu}$ and a closed two-form~$\CF_{\mu\nu}$, as well as the supersymmetry currents~$S_{\mu\alpha}, \t S_{\mu\alphadot}$. The corresponding background fields reside in the linearized version of the new-minimal supergravity multiplet, whose bosonic components are the~$R$-symmetry gauge field~$A_\mu^{(R)}$, the linearized metric~$h_{\mu\nu}$,\foot{Here~$h_{\mu\nu}$ is normalized so that~$g_{\mu\nu} = \delta_{\mu\nu} + 2 h_{\mu\nu}$. In the notation of~\ClossetVRA, we have~$\Delta g_{\mu\nu} = 2 h_{\mu\nu}$.} and a two-form gauge field~$B_{\mu\nu}$, whose dual field strength is the conserved vector~$V^\mu = {i \over 2} \ep^{\mu\nu\rho\lambda} \d_\nu B_{\rho\lambda}$. Supersymmetric configurations are determined by setting the gravitinos~$\Psi_{\mu\alpha}, \t \Psi_{\mu\alphadot}$ and their variations to zero, which leads to the Killing spinor equations~\speq. The couplings to supergravity background fields are then given by
\eqn\rsugra{\SL_{\rm SUGRA} = -h^{\mu\nu} T_{\mu\nu} + A^{(R)\mu} j_\mu^{(R)} + {i \over 4} \ep^{\mu\nu\rho\lambda} \CF_{\mu\nu} B_{\mu\nu} + \cdots~.}
As in~\flcurbf, the ellipsis signifies higher-order terms in the supergravity fields, which are required for invariance under diffeomorphisms, as well as local supersymmetry and~$R$-symmetry transformations. The non-linear completion of~\rsugra\ is the rigid limit of matter-coupled new minimal supergravity, which was reviewed in section~2. 

By supersymmetry, the shift of the~$R$-current in~\jmix\ must be accompanied by changes of the other operators in its supermultiplet, which take the form of improvements~\refs{\KomargodskiRB,\DumitrescuIU},
\eqn\improvfourd{\eqalign{
&j_\mu^{(R)}\rightarrow j_\mu^{(R)} + t j_\mu~,\cr
&T_{\mu\nu}\rightarrow T_{\mu\nu} - {t \over 2} \left(\d_\mu\d_\nu -\delta_{\mu\nu}\d^2\right) J~,\cr
&\CF_{\mu\nu}\rightarrow \CF_{\mu\nu} + {3 t\over 2} \left(\d_\mu j_\nu-\d_\nu j_\mu\right)~,
}}
and similarly for~$S_{\mu\alpha}, \t S_{\mu\alphadot}$. Substituting into~\rsugra\ and comparing with~\flcurbf, we find that these improvements lead to the following changes in the background fields,
\eqn\linbfshift{\Delta a_\mu = t \Big(A_\mu^{(R)} + {3 \over 2} V_\mu\Big)~, \qquad \Delta D = -{t \over 4} R + \CO(V^2)~,}
where~$R = 2 \left(\d^2 {h^\mu}_\mu - \d^\mu \d^\nu h_{\mu\nu}\right) + 
\CO(h^2)$ is the Ricci scalar. Therefore, the flavor gauge field~$a_\mu$ and its~$D$-term are shifted by a certain vector multiplet constructed out of the supergravity fields.\foot{This vector multiplet already played an important role in~\SohniusFW.} Immediately worrisome is the fact that~$A_\mu^{(R)}$ is a gauge field, while~$\Delta a_\mu$ should be a well-defined one-form. We will return to this important point below. Since~\linbfshift\ was derived at the linear level, it may receive higher-order contributions in the supergravity fields. The only such correction compatible with gauge invariance and dimensional analysis is a possible~$V^\mu V_\mu$ term in~$\Delta D$. 

We will now show that such a term is indeed present, and that its coefficient is fixed by supersymmetry. To see this, note that supersymmetry requires the shifts in~\linbfshift\ to satisfy the integrability conditions~\holvecbg\ for background gauge fields, 
\eqn\shiftint{
\d_{\b i} \big(\Delta a_{\b j}\big) - \d_{\b j} \big(\Delta a_{\b i}\big) = 0~, \qquad \Delta D = - i g^{i \b j} \left(\d_i \big(\Delta a_{\b j}\big) - \d_{\b j} \big(\Delta a_i\big)\right)~.
}
The first equation follows from substituting~\fdbg\ into~\linbfshift, while the second equation is satisfied for a unique choice of~$V^\mu V_\mu$ term in~$\Delta D$, 
\eqn\ddnl{
\Delta D = -{t \over 4} \left(R- 6 V^\mu V_\mu \right)~.
}
To check this, it is convenient to use the integrability conditions that follow from evaluating the right-hand side of the identity~$\half R_{\mu\nu\kappa\lambda} \sigma^{\kappa\lambda} \zeta = [\grad_\mu, \grad_\nu] \zeta$ using the Killing spinor equation~\speq.\foot{The explicit form of these integrability conditions is written out in equation~(5.1) of~\DumitrescuHA.} 

Below, we will use the shifts in~\linbfshift\ and~\ddnl\ to determine the dependence of~$Z_\CM$ on the~$R$-symmetry mixing parameter~$t$. It is therefore instructive to re-derive these shifts using twisted variables. For simplicity, we consider a single twisted chiral superfield of~$R$-charge~$r$ and~$U(1)$ flavor charge~$q$ (and its conjugate twisted anti-chiral superfield) with Lagrangian~\lagform. Shifting the~$R$-symmetry as in~\jmix\ amounts to~$r \rightarrow r + t q$. It follows from~\chimul\ and~\formsach\ that the component fields of the twisted chiral and anti-chiral multiplets change as follows,
\eqn\tfch{
\left(\CC, \CX, \CX_{ij}, \CM_{ij}\right) \rightarrow p^{-{tq\over 2}} \left(\CC, \CX, \CX_{ij}, \CM_{ij}\right)~, \quad \big(\t \CC, \t \CX_{\b i}, \t \CM_{\b i \b j}\big) \rightarrow p^{t q \over 2} \big(\t \CC, \t \CX_{\b i}, \t  \CM_{\b i \b j}\big)~.
}
Substituting into the Lagrangian~\lagform\ and evaluating the various covariant derivatives with the help of formulas in appendix B, we find that the changes in~\tfch\ lead to the following shifts of the background flavor gauge field~$a_\mu$ and its twisted superpartner~$\CD$,
\eqn\twgfshift{\eqalign{
& \Delta a_i = t \left({i \over 8} \d_i \log g -{i \over 2} \d_i \log s + V_i + \half U_i\right)~,\cr
& \Delta a_{\b i} = t \left(-{i \over 8} \d_{\b i} \log g -{i \over 2} \d_{\b i} \log s\right)~,\cr
& \Delta \CD = 0~.
}}
Using~\fdbg\ and~\fdbgii, the first two equations reduce to~$\Delta a_\mu = t \left(A_\mu^{(R)} + {3 \over 2} V_\mu\right)$, in agreement with~\linbfshift. The equivalence of~$\Delta \CD = 0$ and~\ddnl\ has already been established above.

\subsec{Consequences for the Partition Function}

To summarize, we have found that the shift~\jmix\ of the~$R$-symmetry changes the background flavor gauge field~$a_\mu$ and its superpartners according to~\linbfshift\ and~\ddnl, or equivalently~\twgfshift. In terms of the flavor line bundle~$\CL$ and the~$R$-symmetry line bundle~$L$, this amounts to shifting~$\CL\rightarrow \CL \otimes L^t$. Continuous variations of~$t$ are therefore  only possible if~$L$ is topologically trivial, $c_1(L)=0$, since the topology of~$\CL$ cannot change continuously. In this case~$A_\mu^{(R )}$ can essentially be treated like a well-defined one-form, so that the expression for~$\Delta a_\mu$ in~\linbfshift\ is meaningful. As was shown in~\DumitrescuHA\ and reviewed in section~3.1, the holomorphic twist required to place an~$\CN=1$ theory on the complex manifold~$\CM$ identifies~$L = \CK^{-\half}$, where~$\CK$ is the canonical bundle of~$\CM$. Therefore~$t$ is unrestricted as long as~$c_1(\CK) = 0$, so that the canonical bundle of~$\CM$ is topologically trivial.\foot{When~$c_1(\CK) \neq 0$, there are quantization conditions on the allowed~$R$-charges, and hence on~$t$. In particular, we cannot vary~$t$ continuously. See~\ClossetVRA\ and references therein for further discussion.} 

This notwithstanding, $\CK$ may be non-trivial as a holomorphic line bundle, so that the shift~$\CL \rightarrow \CL \otimes \CK^{-{t \over 2}}$ leads to a~$t$-dependent change in the holomorphic moduli of the flavor line bundle~$\CL$. As we showed in section~4.2, the partition function~$Z_\CM$ is a holomorphic function of these moduli, and hence it can acquire a dependence on~$t$. More precisely, $Z_\CM$ only depends on the cohomology class of~$\Delta a_{\b i}$ in~$H^{0,1}(\CM)$, with~$\Delta a_{\b i}$ given by~\twgfshift, while~$\Delta a_{i}$ and~$\Delta D$ do not affect~$Z_\CM$.

To make this explicit, we choose a basis~$\omega_A^{0,1}$ for~$H^{0,1}(\CM)$, $A = 1, \ldots, \dim H^{0,1}(\CM)$. Once we have picked a fiducial flavor line bundle~$\CL$ of fixed topology, we can parametrize all other holomorphic line bundles of the same topology by specifying an element~$\nu^A \omega_A^{0,1} \in H^{0,1}(\CM)$. Up to global identifications, the constants~$\nu^A \in \C$ are coordinates on the moduli space of holomorphic line bundles whose topology coincides with the topology of~$\CL$. In these coordinates, the shift~$\CL \rightarrow \CL \otimes \CK^{-{t \over 2}}$ is described by
\eqn\deltanu{
\nu^A \rightarrow \nu^A -{t \over 2} \nu^A_\CK~,
}
where the constants~$\nu^A_\CK$ are determined by comparing~\twgfshift\ and~\deltanu,
\eqn\canlbcoord{
i \b \d \log p = \nu^A_\CK \omega^{0,1}_{A} \qquad {\rm in} \qquad H^{0,1}(\CM)~.
}
Here we have used~\locdefs\ to express~${i \over 4} \b \d \log g + i \b \d \log s = i \b \d \log p$, where~$p$ is the nowhere vanishing section that trivializes the line bundle~$L^2 \otimes \CK$. 

It follows from~\deltanu\ that the partition function~$Z_\CM$ only depends on the~$R$-symmetry mixing parameter~$t$ through its dependence on holomorphic line bundle moduli,
\eqn\zrdep{
Z_\CM\left(t, \nu^A\right) = Z_\CM\Big(0, \nu^A -{t \over 2} \nu^A_\CK\Big)~.
}
Several comments are in order:
\medskip
\item{1.)} Since~$Z_\CM(0, \nu^A)$ is a locally holomorphic function of the line bundle moduli~$\nu^A$, $Z_\CM(t, \nu^A)$ is locally holomorphic in the complex linear combinations~$\nu^A - {t \over 2} \nu^A_\CK$. 
\medskip
\item{2.)} The mixing parameter~$t$ as defined in~\jmix\ is real. It is natural to continue it to complex values, in which case~$Z_\CM(t, \nu^A)$ is locally holomorphic in~$t$.
\medskip
\item{3.)} The~$\nu^A_\CK$, which are determined by~\canlbcoord, are coordinates that specify the location of the canonical bundle within the moduli space of of holomorphic line bundles of trivial topology. Consequently, they depend holomorphically on the complex structure moduli of~$\CM$, but not the metric. Any apparent metric dependence in~\canlbcoord\ vanishes in cohomology. This is required for consistency with the results of section~4.4, according to which~$Z_\CM(t, \nu^A)$ is independent of the Hermitian metric and locally holomorphic in the complex structure moduli of~$\CM$ for all values of~$t$. 
\medskip
\item{4.)} If~$\CK$ is trivial as a holomorphic line bundle, all~$\nu_\CK^A$ vanish. In this case the partition function does not depend on the choice of~$R$-symmetry. This is consistent with the fact that complex manifolds with holomorphically trivial canonical bundle can serve as supersymmetric backgrounds for field theories that do not possess an~$R$-symmetry~\DumitrescuAT. 

\medskip

\subsec{Example: $S^3\times S^1$ and the Supersymmetric Index}

Here we apply~\zrdep\ to analyze the~$R$-symmetry dependence of the partition function on complex manifolds~$\CM$ that are diffeomorphic to~$S^3 \times S^1$. Such complex manifolds, known as primary Hopf surfaces, were one of the main examples studied in~\ClossetVRA, which contains further details and references. (See also the recent discussion in~\AsselPAA.)

A primary Hopf surface~$\CM^{p,q}$ is defined by the following quotient of~$\C^2 -(0,0)$:
\eqn\Hofpsurf{
\left(w,z\right) \sim \left(p w, q z\right)~, \qquad 0 < |p| \leq |q| <1~.
}
Here~$p,q$ are the two complex structure moduli of~$\CM^{p,q}$, which we will express as follows,\foot{In the notation of~\ClossetVRA, we have~$\sigma = {\vartheta_p \over 2 \pi} + {i \beta_p \over 2 \pi}$ and~$\tau = {\vartheta_q \over 2 \pi} +{i \beta_q \over 2 \pi}$~.}
\eqn\pqdef{\eqalign{
& p= e^{2\pi i \sigma}~, \qquad \Re \sigma \sim \Re \sigma + 1~,\cr
& q= e^{2\pi i \tau}~, \qquad \Re \tau \sim \Re \tau +1~.
}}
We can trivialize the identifications in~\Hofpsurf\ by introducing real coordinates~$x, \theta, \varphi, \chi$,
\eqn\zinangles{\eqalign{
& w = e^{2\pi i \sigma x}\cos{\theta\over 2} e^{i\varphi}~, \qquad
z = e^{2 \pi i\tau x}\sin{\theta\over 2} e^{i\chi}~,\cr
& x \sim x+1~, \quad 0 \leq \theta \leq \pi~, \quad \varphi \sim \varphi + 2\pi~, \quad \chi \sim \chi+2\pi~.
}}
This is an explicit diffeomorphism between~$\CM^{p,q}$ and~$S^3 \times S^1$, with~$x$ running along~$S^1$ and~$\theta, \varphi, \chi$ parametrizing~$S^3$. 

All complex line bundles over~$\CM^{p,q}$, including the canonical bundle~$\CK$, are topologically trivial, and hence the results of section~5.2 apply. The Dolbeault cohomology~$H^{0,1}(\CM)$ is one-dimensional, so that all holomorphic line bundles possess a single modulus. We will choose the following basis element for~$H^{0,1}(\CM)$,
\eqn\defomega{
\omega^{0,1}= \b\d(-2x)~.
}
The~$(0,1)$-part of a supersymmetric Abelian flavor gauge field~$a_\mu$ can be written as\foot{Our normalization of~$\nu$ is such that~$\nu = - \half\left({a_r - i a_i}\right)$ in the notation of~\ClossetVRA.}
\eqn\abackgdgen{
\left(a_\mu dx^\mu\right)^{0,1}= \nu\omega^{0,1}~, \qquad \nu \in \C~.
}

In order to apply~\canlbcoord\ and evaluate~$\nu_\CK$, we must determine the section~$p$. Demanding that~$p dw \wedge dz$ is well defined and nowhere vanishing, we find that
\eqn\holtwo{p= s_0 e^{-2\pi i \left(\sigma + \tau\right) x}~,}
where~$s_0$ is a well-defined, nowhere vanishing complex function on~$\CM^{p,q}$, which must be invariant under the identifications in~\Hofpsurf. Note that fixing the ambiguity of~$\sigma, \tau$ in~\pqdef\ amounts to choosing a homotopy class for~$s_0$. Substituting~\holtwo\ into~\canlbcoord, we find
\eqn\nucalc{
\nu_\CK = - \pi \left(\sigma + \tau\right)~.
}
Further substituting into~\deltanu, we obtain the following shift in~$\nu$,
\eqn\shiftRhopf{
\nu \rightarrow \nu + {\pi t\over 2} \left(\sigma+\tau\right)~.
}
Note that the properties of~$\nu_\CK$ in~\nucalc\ are consistent with the general comments at the end of section~5.2: it depends holomorphically on the complex structure moduli~$\sigma$ and~$\tau$, but it is independent of the metric. (In fact, at no point did we need to specify a metric on~$\CM^{p,q}$.) Moreover, $\nu_\CK$ is a well-defined number once we fix the ambiguity in~$\Re \sigma$ and~$\Re \tau$ by choosing a homotopy class for~$s_0$. 

As explained in~\ClossetVRA, the partition function~$Z_{\CM^{p,q}}(\nu)$ coincides with the supersymmetric index computed on~$S^3 \times \R$, up to local counterterms and possible quantum anomalies,
\eqn\defIndex{\CI(p,q, u) = \Tr_{S^3} \left((-1)^F p^{J_3 + J'_3 - {R \over 2}} q^{J_3 - J'_3 - {R \over 2}} u^{Q_f} \right)~.}
Here~$F$ is the fermion number. The symmetry of the theory on~$S^3 \times \R$ is~$SU(2|1) \times SU(2)'$, with~$J_3$ and~$J_3'$ the Cartan generators of~$SU(2) \subset SU(2|1)$ and~$SU(2)'$, respectively. The~$R$-charge is denoted by~$R$, and~$Q_f$ is an Abelian flavor charge. The fugacity~$u$ conjugate to~$Q_f$ is related to the holomorphic line bundle modulus~$\nu$ in~\abackgdgen\ as  follows (see~\ClossetVRA\ for details),
\eqn\defu{
u= e^{-2 i \nu}~.
}
If we shift~$R\rightarrow R + t Q_f$ in the definition~\defIndex\ of the index, we find that
\eqn\shiftupq{
u\rightarrow u \left(pq\right)^{-{t\over 2}}~,
}
in perfect agreement with~\shiftRhopf\ and~\defu.

\newsec{Generalization to Three Dimensions}

The bulk of this paper has been dedicated to four-dimensional~$\CN=1$ theories with a~$U(1)_R$ symmetry on complex manifolds and their description using twisted variables. In this section, we sketch the necessary ingredients to generalize these results to three-dimensional~$\CN=2$ theories with a~$U(1)_R$ symmetry on curved manifolds~$\CM$. As in flat space, they closely resemble their four-dimensional counterparts. The underlying geometric structure is a transversely holomorphic foliation (THF), which endows the three-manifold~$\CM$ with a near-perfect analogue of complex geometry~\refs{\ClossetRU,\ClossetVRA}. (See \KlareGN\ for related earlier work.)
 Below, we briefly review basic aspects of THFs in three dimensions and indicate how they can be used to generalize the results of sections~3 and~4. Following the discussion in section~5, we analyze the~$R$-symmetry dependence of supersymmetric partition functions on~$\CM$. In particular, we will obtain an explicit formula for general squashed spheres.

\subsec{Killing Spinors and Transversely Holomorphic Foliations}

In this section we closely follow the discussion in~\refs{\ClossetRU,\ClossetVRA}, to which we refer for further reading.\foot{See also~\refs{\DKi\GM-\Haef} for a general discussion of THFs and~\refs{\BG\Brunella-\Ghys} for results in three dimensions.} The bosonic supergravity fields that describe supersymmetric backgrounds for three-dimensional~$\CN=2$ theories with a~$U(1)_R$ symmetry are the metric~$g_{\mu\nu}$, two Abelian gauge fields~$A_\mu^{(R)}$ and~$C_\mu$, which couple to the~$R$-symmetry and the central charge, and a scalar~$H$. The dual field strength~$V^\mu = - i \ep^{\mu\nu\rho} \d_\nu C_\rho$ is a covariantly conserved vector. The supergravity multiplet also contains complex gravitinos~$\Psi_{\mu\alpha}, \t \Psi_{\mu\alpha}$. Supersymmetric backgrounds\foot{See~\KuzenkoUYA\ for a recent discussion of the full supergravity theory in components.} are determined by setting the gravitinos and their supersymmetry variations to zero, which leads to the following Killing spinor equation,
\eqn\tdkseq{
\big(\grad_\mu - i A_\mu^{(R)}\big)\zeta = - \half H \gamma_\mu \zeta + {i \over 2} V_\mu \zeta - \half \ep_{\mu\nu\rho} V^\nu \gamma^\rho \zeta~,
}
and a similar equation for the conjugate spinor~$\t \zeta_\alpha$, which will not be needed here. A Killing spinor~$\zeta_\alpha$ satisfying~\tdkseq\ exists if and only if the three-manifold~$\CM$ admits a THF and~$g_{\mu\nu}$ is a compatible transversely Hermitian metric.

On a three-manifold~$\CM$, a THF consists of a one-dimensional oriented foliation and an integrable complex structure~$J$ on the two-dimensional normal bundle~$\CD$ of the foliation. It is always possible to find a nowhere vanishing vector field~$\xi^\mu$, whose orbits are the leaves of the foliation. Given a THF, we can cover~$\CM$ with patches of adapted coordinates ~$\tau, z, \b z$. The real coordinate~$\tau$ parametrizes the leaves of the foliation, so that~$\xi = \d_\tau$, while~$z$ is a holomorphic coordinate on the normal bundle~$\CD$. Two overlapping adapted coordinate systems are related by
\eqn\holcc{
\tau' = \tau + t(z, \b z)~, \qquad z' = f(z)~, 
}
where~$t(z, \b z)$ is real and~$f(z)$ is holomorphic. In adapted coordinates, a transversely Hermitian metric takes the following form:
\eqn\cthm{
ds^2 = \left(d\tau + h dz + \b h d \b z\right)^2 + c^2 dz d \b z~,
}
where~$h(\tau, z, \b z)$ is complex and~$c(\tau, z, \b z)$ is real. Note that~$\xi = \d_\tau$ is generally not a Killing vector. We can use such a metric to define two useful auxiliary objects,
\eqn\etaphidef{
\eta_\mu dx^\mu = g_{\mu\nu} \xi^\nu dx^\mu = d\tau + h dz + \b h d\b z~, \quad {\Phi^\mu}_\nu = - {\ep^\mu}_{\nu\rho} \xi^\rho = \pmatrix{0 & - i h & i \b h \cr 0 & i & 0 \cr 0 & 0 & -i}~.
}
Note that the projection of~${\Phi^\mu}_\nu$ onto its~$z, \b z$ components is nothing but the complex structure of the normal bundle, $\Phi|_\CD = J$.

The properties of~THFs, and in particular the form of the coordinate transformations~\holcc, allow the definition of many structures that are familiar from complex geometry. For instance, we can split the bundle of complex differential forms into~$(p,q)$-forms and define an analogue of the~$\b \d$-operator, $\t \d$, which maps~$(p,q)$-forms into~$(p,q+1)$-forms and satisfies~$\t \d^2 = 0$ (see~\ClossetVRA\ and references therein). For instance, $\omega^{1,0} = \omega^{1,0}_z dz$ is a~$(1,0)$-form. Under an adapted coordinate change~\holcc, it transforms as follows,
\eqn\cbholtrans{
\left(\omega'^{1,0}\right)_{z'} = {1 \over f'(z)} \omega_z^{1,0}~.
}
In general, we refer to a line bundle whose transition functions are independent of~$\tau$ and holomorphic in~$z$ as a holomorphic line bundle over~$\CM$. (As we will review below, supersymmetric configurations for Abelian background gauge fields correspond to holomorphic line bundles.) It follows from~\cbholtrans\ that the complex~$(1,0)$-forms constitute a holomorphic line bundle, which we refer to as the canonical bundle~$\CK$ of the THF. As in complex geometry, holomorphic line bundles form moduli spaces, which are locally described by the Dolbeault cohomology~$H^{0,1}(\CM)$ (now defined using the~$\t \d$-operator), whose dimension counts the number of moduli.  Similarly, the THF itself typically belongs to a complex moduli space. THFs on compact three-manifolds have been classified in~\refs{\BG\Brunella-\Ghys}.

Every non-trivial solution~$\zeta_\alpha$ of the Killing spinor equation~\tdkseq\ is everywhere non-zero. The relation between~$\zeta_\alpha$ and the THF is then given by the following formula:
\eqn\etazetarel{
\eta_\mu = {1 \over |\zeta|^2} \zeta^\dagger \gamma_\mu \zeta~.
}
We can also use~$\zeta_\alpha$ to construct a nowhere vanishing section~$P_\mu = \zeta \gamma_\mu \zeta$, which trivializes the bundle~$L^2 \otimes \CK$. Here~$L$ is the~$U(1)_R$ line bundle and~$\CK$ is the canonical bundle of the THF defined by~$\zeta_\alpha$. As in four dimensions, this defines a holomorphic twist that identifies~$L = \CK^{-\half}$. After the twist, the supercharge~$Q$ corresponding to~$\zeta_\alpha$ transforms as a scalar under adapted coordinate changes~\holcc. The Killing spinor equation~\tdkseq\ then essentially determines the supergravity background fields in terms of geometric data,
\eqn\tdrgf{\eqalign{
&V^\mu = \ep^{\mu\nu\rho} \d_\nu \eta_\rho + U^\mu + \kappa \eta^\mu~, \cr
& H = - \half \grad_\mu \eta^\mu + {i \over 2} \ep^{\mu\nu\rho} \eta_\mu \d_\nu \eta_\rho + i \kappa~,\cr
& {\Phi^\mu}_\nu U^\nu = -i U^\mu~, \qquad \grad_\mu \left(U^\mu + \kappa \eta^\mu\right) = 0~,}}
and
\eqn\tdrgf{\eqalign{
&A^{( R)}_\mu =  \hat A_\mu + A_\mu^{\rm flat} -\half  \ep_{\mu\nu\rho} \d^\nu \eta^\rho + {i \over 4} \eta_\mu \grad_\nu \eta^\nu - {i \over 2} \eta^\nu \grad_\nu \eta_\mu~,\cr
&  \hat A_\mu = {1 \over 8} {\Phi^\nu}_\mu \d_\nu \log g~, \qquad A_\mu^{\rm flat} = -{i\over 2}\d_\mu\log s~.}}
The formula for~$\hat A_\mu$ is only valid in coordinates adapted to the THF, while the flat connection~$A_\mu^{\rm flat}$ is determined by~$s = p g^{-{1 \over 4}}$ with~$p = P_z$ and~$g = \det (g_{\mu\nu})$.  The comments at the end of section~3.1 also apply here. 

By comparing the preceding discussion with the four-dimensional setup reviewed in section~3.1, it is clear that the geometric ingredients are essentially identical. It is therefore straightforward (but tedious) to start with the untwisted fields and Lagrangians described in~\ClossetRU\ and introduce twisted variables adapted to a choice of THF on~$\CM$. Similarly, we could follow the logic of section~4 and use this twisted description to study the dependence of the supersymmetric partition function~$Z_\CM$ on the background geometry. As in four dimensions, this would lead to an alternative derivation of the results obtained in~\ClossetVRA: the partition function is a locally holomorphic function of the complex moduli that parametrize the THF and holomorphic line bundles. However, $Z_\CM$ does not depend on the choice of transversely Hermitian metric.

\subsec{Dependence of the Partition Function on the~$R$-Symmetry}

We will now generalize the results of section~5, which did not require the machinery of twisted variables,  to three dimensions. As in~\jmix, we consider a~$t$-dependent shift of the~$R$-current~$j_\mu^{(R)}$ by an Abelian flavor current~$j_\mu$,
\eqn\jmixii{
j_\mu^{(R)} \rightarrow j_\mu^{(R)} + t j_\mu~.
}
The current~$j_\mu$ resides in a real linear multiplet with real scalars~$J, K$ and fermions~$j_\alpha, \t j_\alpha$. These operators couple to a background gauge field~$a_\mu$, scalars~$D, \sigma$ and gauginos~$\lambda_\alpha, \t \lambda_\alpha$,
\eqn\tdlcurbf{
{\scr L}_{\rm flavor} = a^\mu j_\mu + \sigma K + D J + \lambda^\alpha j_\alpha - \t \lambda^\alpha \t j_\alpha + \cdots~.
}
Supersymmetric configurations are determined by setting~$\lambda_\alpha, \t \lambda_\alpha$ and their supersymmetry variations to zero. The resulting constraints on the background fields are conveniently expressed by introducing a complex gauge field~$\SA_\mu$ and its field strength,
\eqn\sadef{\SA_\mu = a_\mu + i \sigma \eta_\mu~,\qquad {\scr F}_{\mu\nu} = \d_\mu \SA_\nu - \d_\nu \SA_\mu~.}
In adapted coordinates, these constraints take the form
\eqn\tdgfcons{{\scr F}_{\tau \b z} = 0~,\qquad D = - \half \Phi^{\mu\nu} \SF_{\mu\nu} + \eta^\mu \d_\mu \sigma + \sigma \left(\half \grad_\mu \eta^\mu - {i \over 2} \ep^{\mu\nu\rho} \eta_\mu \d_\nu \eta_\rho\right)~.}
It was shown in~\ClossetVRA\ that the first equation defines a holomorphic line bundle over~$\CM$, just as in four dimensions. 

The~$R$-current~$j_\mu^{(R)}$ resides in the same supermultiplet as the supersymmetry currents~$S_{\mu\alpha}, \t S_{\mu\alpha}$, the energy-momentum tensor~$T_{\mu\nu}$, the central charge current~$j_\mu^{(Z)}$, and a real scalar~$J^{(Z)}$.  Their linearized couplings to the bosonic supergravity fields introduced in section~6.1 (with~$g_{\mu\nu} = \delta_{\mu\nu} + 2 h_{\mu\nu}$) are given by
\eqn\tdsugracoup{
{\scr L}_{\rm SUGRA} = - h^{\mu\nu} T_{\mu\nu} + A^{(R) \mu} j_\mu^{(R)} + C^\mu j_\mu^{(Z)} + H J^{(Z)} + \cdots. 
}
As in four dimensions, the shift~\jmixii\ of the~$R$-current induces improvement transformations of the other operators in its supermultiplet. For the bosonic operators, we find~\refs{\DumitrescuIU,\ClossetVP}
\eqn\trmimps{\eqalign{
& j_\mu^{(R)} \rightarrow j_\mu^{(R)} + t j_\mu~,\cr
& T_{\mu\nu} \rightarrow T_{\mu\nu} -{t \over 2} \left(\d_\mu \d_\nu - \delta_{\mu\nu} \d^2 \right) J~,\cr
& J^{(Z)} \rightarrow J^{(Z)} + t K~,\cr
& j_\mu^{(Z)} \rightarrow j_\mu^{(Z)} - i t \ep_{\mu\nu\rho} \d^\nu j^\rho~.
}}
Substituting into~\tdsugracoup\ and comparing with~\tdlcurbf, we find the following linearized shifts:
\eqn\tdlinshift{
\Delta a_\mu = t\big(A_\mu^{(R)} + V_\mu\big)~, \qquad \Delta \sigma = t H~, \qquad \Delta D = -{t \over 4} \left(R + \cdots\right)~,
}
where the ellipsis in~$\Delta D$ denotes possible higher-order terms in the supergravity fields that are allowed by dimensional analysis and gauge invariance. As before, they can be fixed by demanding that the shifts in~\tdlinshift\ satisfy the constraints~\tdgfcons. A calculation using the integrability conditions\foot{These integrability conditions are written out in equation~(5.8) of~\ClossetRU.} that follow from the Killing spinor equation~\tdkseq\ shows that
\eqn\tdnldeltad{
\Delta D = -{t \over 4} \left(R - 2 V^\mu V_\mu - 2 H^2 \right)~.
}

We can now repeat the arguments in section~5 to deduce the~$t$-dependence of the partition function~$Z_\CM$. As was the case there, continuous shifts of~$t$ require the canonical bundle~$\CK$ of the THF to be topologically trivial, $c_1(\CK) = 0$, which renders the expression for~$\Delta a_\mu$ in~\tdlinshift\ sufficiently well defined. Recall the following result of~\ClossetVRA:  the deformation of the flavor line bundle induced by the shifts~\tdlinshift\ and~\tdnldeltad\ is parametrized by the following one-form, appropriately projected onto cohomology,
\eqn\deltasc{
\Delta \SA^{0,1} = \left(\Delta a_\mu dx^\mu + i \Delta \sigma \eta_\mu dx^\mu\right)^{0,1} \qquad {\rm in} \qquad H^{0,1}(\CM)~.
}
If we choose a basis~$\omega_A^{0,1}$ for~$H^{0,1}(\CM)$, the moduli space for holomorphic line bundles of fixed topology is parametrized by linear combinations~$\nu^A \omega^{0,1}_A \in H^{0,1}(\CM)$. The shift in~\deltasc\ then takes the form
\eqn\tdnushift{
\nu^A \rightarrow \nu^A -{t \over 2} \nu_\CK^A~,
}
where the constants~$\nu_\CK^A$ are determined by solving
\eqn\tdnuksolve{
i \t \d \log p = \nu_\CK^A \omega_A^{0,1} \qquad {\rm in} \qquad H^{0,1}(\CM)~.
}
Here we have substituted~\tdrgf\ and \tdlinshift\ into~\deltasc\ and used the fact that the trivializing section of~$L^2 \otimes \CK$ is given by~$p = g^{1 \over 4} s$. Note that~\tdnuksolve\ takes the same form as~\canlbcoord\ if we replace the~$\b \d$-operator by its three-dimensional counterpart~$\t \d$. Since the partition function is locally holomorphic in the line bundle moduli~$\nu^A$, its dependence on the~$R$-symmetry mixing parameter~$t$ is completely determined by~\tdnushift, exactly as in~\zrdep,
\eqn\zrdep{
Z_\CM\left(t, \nu^A\right) = Z_\CM\Big(0, \nu^A -{t \over 2} \nu^A_\CK\Big)~.
}
The comments at the end of section~5.2 are straightforwardly adapted to the present case.

\subsec{Example: Squashed Spheres}

We will illustrate the results of the previous subsection by providing an {\it a priori} explanation for the~$R$-symmetry dependence of supersymmetric partition functions on squashed three-spheres, i.e.~manifolds diffeomorphic but not necessarily isometric to a round~$S^3$. As was shown in~\ClossetVRA, most known examples of supersymmetric squashed three-spheres belong to a one-parameter family of THFs on~$S^3$. (An example that does not belong to this family was also constructed there.) In concrete calculations, we can choose any convenient representative of this one-parameter family. Here we will use the~$U(1)\times U(1)$-symmetric squashed sphere~$S^3_b$ of~\HamaEA. For these backgrounds, the squashing parameter~$b$, which corresponds to the THF modulus, is real. Since the partition function~$Z_{S^3_b}$ is locally holomorphic in~$b$, we can analytically continue to complex~$b$ at the end of the calculation~\ClossetVRA. 

In our conventions, the supergravity background fields that describe the squashed sphere of~\HamaEA\ take the following form:
\eqn\HHLback{\eqalign{
&ds^2= {1 \over 4} f(\theta)^2 d\theta^2 + b^{-2} \cos^2{\theta\over 2}d\varphi^2 + b^2\sin^2{\theta\over 2}d\chi^2~,\cr
& V^\mu = 0~, \qquad H = {i\over f(\theta)}~, \cr
&  A^{(R)}_\mu dx^\mu =\half \left(1-{1\over b f(\theta)}\right) d\varphi + \half\left(1-{b\over  f(\theta)}\right) d\chi~,
}}
while the vector field whose orbits determine the leaves of the THF is given by 
\eqn\THFHHL{
\eta^\mu \d_\mu = - b\d_\varphi -b^{-1}\d_\chi~.
}
Here~$0 \leq \theta \leq \pi$, $\varphi \sim \varphi + 2\pi$, and~$\chi \sim \chi+2\pi$. The function~$f(\theta)$ was chosen to have a specific form in~\HamaEA, but it was shown in~\MartelliFU\ that~$f(\theta)$ is essentially arbitrary (up to to smooth boundary conditions at~$\theta = 0,\pi$) and does not affect the partition function~$Z_{S^3_b}$. In the language of~\ClossetVRA, different choices of~$f(\theta)$ parametrize distinct transversely Hermitian metrics for the same THF, and hence these choices do not affect~$Z_{S^3_b}$. Also note that we have set the overall dimensionful `radius' of the metric to one. 

Since all complex line bundles over~$S^3_b$ are topologically trivial, we can apply the results of section~6.2. The Dolbeault cohomology~$H^{0,1}(S^3_b)$ was explicitly constructed in~\ClossetVRA. It is one-dimensional and generated by~$\omega^{0,1}_\mu = \eta_\mu$. Therefore, we can parametrize supersymmetric Abelian flavor gauge fields on~$S^3_b$ as follows,
\eqn\tdsfgf{
\SA_\mu = \nu \omega_\mu^{0,1} = \nu\eta_\mu~, \qquad \nu \in \C~.
}
Here~$\SA_\mu$ is the complex gauge field defined in~\sadef. Instead of using~\tdnuksolve\ to evaluate~$\nu_\CK$, we can simply substitute the explicit supergravity background fields~\HHLback\ into~\tdlinshift\ and~\deltasc. After projecting onto~$\eta_\mu$, we find that
\eqn\tdnucalc{
\nu_\CK = b + b^{-1}~.
}
As expected, this only depends on the THF modulus~$b$. All extraneous data, such as the function~$f(\theta)$ that appears in~\HHLback, have dropped out. Substituting~\tdnucalc\ into~\tdnushift, we obtain the change in~$\nu$ under a~$t$-dependent shift of the~$R$-symmetry,
\eqn\tdnushiftex{
\nu \rightarrow \nu -{t \over 2} \left(b + b^{-1}\right)~.
}
Therefore, $Z_{S^3_b}$ only depends on~$\nu$ and~$t$ through a holomorphic function of~$\nu -{t \over 2} \left(b + b^{-1}\right)$. 

As was explained in~\ClossetRU, turning on a real mass~$m$ for the flavor current~$j_\mu$ amounts to setting~$\nu = i m$. Therefore, the partition function on~$S^3_b$ in the presence of the real mass~$m$ is a locally holomorphic function of
\eqn\mystholo{
m +{i t\over 2} \left(b+ b^{-1}\right)~,
}
in perfect agreement with the results of~\HamaEA. A round sphere preserving four supercharges~\refs{\KapustinKZ,\JafferisUN,\HamaAV} corresponds to~$b = 1$. In this case, the partition function is holomorphic in~$m + i t$. The fact that the~$R$-symmetry mixing parameter~$t$ acts as an imaginary part for the real mass played a crucial role in the proof of the~$F$-maximization principle~\refs{\JafferisUN,\ClossetVG}, which determines the superconformal~$R$-symmetry in three-dimensional~$\CN=2$ theories. This holomorphy property was recognized in~\JafferisUN\ as an empirical feature of the explicit partition functions computed in~\refs{\JafferisUN,\HamaAV}, and further explored in~\refs{\FestucciaWS, \BeniniMF}. Our general proof emphasizes the geometric origin of the various parameters in~\mystholo, which enabled us to apply the results of~\ClossetVRA\ about the dependence of~$Z_\CM$ on the background geometry. 

\vskip1.4cm

\noindent {\bf Acknowledgments:}
We would like to thank~M.~Kontsevich, N.~Seiberg, and~E.~Witten for useful discussions. We are grateful to the Kavli Institute for Theoretical Physics, Santa Barbara, where some of this research was carried out, for its hospitality, supported in part by NSF grant PHY11-25915. TD is supported by the Fundamental Laws Initiative of the Center for the Fundamental Laws of Nature at Harvard University, as well as DOE grant~DE-SC0007870 and NSF grants~PHY-0847457, PHY-1067976. GF is supported by a Mobilex grant from the Danish Council for Independent Research. ZK thanks the Perimeter Institute for Theoretical Physics, Waterloo, for its kind hospitality during the final stages of this project. Research at the Perimeter Institute is supported in part by the Government of Canada through NSERC and by the Province of Ontario through MRI. ZK is supported by the ERC STG grant 335182, by the Israel Science Foundation under grant 884/11, by the United States-Israel Binational Science Foundation (BSF) under grant 2010/629, as well as by the I-CORE Program of the Planning and Budgeting Committee and by the Israel Science Foundation under grant 1937/12. Any opinions, findings, and conclusions or recommendations expressed herein are those of the authors and do not necessarily reflect the views of the funding agencies.

\vskip1.4cm

\appendix{A}{Product Multiplets}

Given two general multiplets~$S_1, S_2$ as in~\genSfourd\ with bottom components~$C_1, C_2$ of~$R$-charge~$r_1, r_2$, we define a product multiplet~$S$ through its bottom component~$C = C_1 C_2$. Using the supersymmetry transformations~\fdgmul, we can then obtain multiplication rules for all other components of~$S$,

\eqn\fdprodmul{\eqalign{& C = C_1 C_2~,\cr
& \chi = \chi_1 C_2 + C_1 \chi_2~,\qquad 
 \t \chi =  \t \chi_1 C_2 + C_1 \t \chi_2~,\cr
& M = M_1 C_2 + C_1 M_2 - i \chi_1 \chi_2~,\qquad
 \t M = \t M_1 C_2 + C_1 \t M_2 + i \t \chi_1 \t \chi_2~,\cr
& a_\mu = a_{1 \mu} C_2 + C_1 a_{2 \mu} + \half \left(\chi_1 \sigma_\mu \t \chi_2 - \t \chi_1 \, \t \sigma_\mu \chi_2\right)~,\cr
& \lambda = \Big(\lambda_1 C_2  + {i \over 2} \t M_1 \chi_2 + {1 \over 2} \sigma^\mu \t \chi_1 \left(a_{2\mu}-iD_\mu C_2\right) \Big)+(1\leftrightarrow 2)~,\cr
& \t \lambda = \Big(\t \lambda_1 C_2  - {i \over 2} M_1 \t \chi_2- {1 \over 2} \t \sigma^\mu \chi_1 \left(a_{2\mu} + i D_\mu C_2\right)\Big)+(1\leftrightarrow 2)~,\cr
& D = D_1 C_2 + C_1 D_2 + \half M_1 \t M_2 + \half \t M_1 M_2 - a_1^\mu a_{2 \mu} - D^\mu C_1 D_\mu C_2 \cr
& \hskip20pt - \chi_1 \Big(\lambda_2 +{i \over 2} \sigma^\mu D_\mu \t \chi_2\Big) - \t \chi_1 \Big(\t \lambda_2 +{i\over 2} \t \sigma^\mu D_\mu \chi_2\Big) - \Big(\lambda_1 -{i \over 2} D_\mu \t \chi_1 \t \sigma^\mu\Big) \chi_2 \cr
& \hskip20pt  - \Big(\t \lambda_1 -{i \over 2} D_\mu \chi_1 \sigma^\mu\Big) \t \chi_2 + {3 \over 2} V_\mu \left(\chi_1 \sigma^\mu \t \chi_2 - \t \chi_1 \t \sigma^\mu \chi_2\right)~.}}
In terms of the twisted variables defined in~\defforms, we can rewrite these multiplication rules as follows, 
\eqn\multrules{\eqalign{&\CC= \CC_1 \CC_2~, \qquad \CX= \CX_1 \CC_2 + \CC_1 \CX_2~,\cr
& \CX_{ij} = \CX_{1, ij} \CC_2 + \CC_1 \CX_{2,ij}~,\qquad \CM_{ij} = \CM_{1,ij} \CC_2 + \CC_1 \CM_{2, ij} + \CX_1 \CX_{2, ij} - \CX_{1, ij} \CX_2~,\cr
& \t \CX_{\b i} = \t \CX_{1, \b i} \CC_2 + \CC_1 \t \CX_{2, \b i}~,\qquad \CA_{\b i} = \CA_{1, \b i} \CC_2 + \CC_1 \CA_{2, \b i} + \CX_1 \t \CX_{2, \b i} - \t \CX_{1, \b i} \CX_2~,\cr
& \CA_i = \CA_{1, i} \CC_2 + \CC_1 \CA_{2, i} + i \CX_{1, ij} \t \CX_2^j - i \t \CX_1^j \CX_{2, ij}~,\cr
& \t \CL_i =\left( \t \CL_{1, i} \CC_2  + \CA_{1, i} \CX_2 - i \CA_1^j \CX_{2, ij} + i \CM_{1, ij} \t \CX_2^j \right) + \left(1 \leftrightarrow 2\right)~,\cr
& \t \CM_{\b i \b j} = \t \CM_{1, \b i \b j} \CC_2 + \CC_1 \t \CM_{2, \b i \b j} + 2i \left(\t \CX_{1, \b i} \t \CX_{2, \b j} - \t \CX_{1, \b j} \t \CX_{2, \b i} \right)~,\cr
& \CL_{\b i \b j} = \left(\CL_{1, \b i \b j} \CC_2  + \t \CM_{1, \b i \b j} \CX_2 + 2i \left(\CA_{1, \b i} \t \CX_{2, \b j} - \CA_{1, \b j} \t \CX_{2, \b i} \right)\right) + \left(1 \leftrightarrow 2\right)~,\cr
& \CL = \left(\CL_1 \CC_2 + {1 \over 4} \t \CM_{1, \b i \b j} \CX_2^{\b i \b j} - \t \CX_1^i \Big(\CA_{2,i} - 2i \grad_i^c \CC_2 - 2 r_2 \big(V_i + \half U_i\big) \CC_2\Big)\right) + \left(1 \leftrightarrow 2\right)~,\cr
& \CD= \bigg(\CD_1 \CC_2 - \CL_1 \CX_2 + {1 \over 4} \CL_{1, \b i \b j} \CX_2^{\b i \b j} +{1 \over 4} \t \CM_{1, \b i \b j} \CM_2^{\b i \b j}   \cr
& \hskip30pt - \CA_1^i \Big(\CA_{2, i} - 2 i \grad_i^c \CC_2 - 2 r_2 \big(V_i + \half U_i\big)\CC_2\Big) \cr
& \hskip30pt + \t\CX_1^i \Big(\t \CL_{2, i} - 2i \grad_i^c \CX_2 -2 r_2 \big(V_i + \half U_i \big) \CX_2\Big) \bigg) + \left(1 \leftrightarrow 2\right)~.
 }}
 \noindent These formulas simplify if we consider products of only twisted chiral multiplets~\chimul\ or only twisted anti-chiral multiplets~\formsach. In this case the multiplication rules~\multrules\ do not explicitly depend on the Hermitian metric or on the vector field~$U^\mu$.

\appendix{B}{Connections on Hermitian Manifolds}

Given a Hermitian manifold with integrable complex structure~${J^\mu}_\nu$ and Hermitian metric~$g_{\mu\nu}$, we would like to define a connection that is compatible with both~${J^\mu}_\nu$ and~$g_{\mu\nu}$. If the manifold is K\"ahler, then~$\grad_\mu {J^\nu}_\rho = 0$ and we can use the usual Levi-Civita connection~$\grad_\mu$. In general, we have to use a connection with torsion. 

A connection~$\hat \grad_\mu$ is metric compatible, i.e.~$\hat \grad_\mu g_{\nu\rho} = 0$, if its connection coefficients~${\, \hat \Gamma^\mu}_{\nu\rho}$ can be expressed as\foot{The spin connection corresponding to~$\hat \grad_\mu$ is given by
\eqn\sc{\hat \omega_{\mu\nu\rho} = \omega_{\mu\nu\rho} -K_{\nu\mu\rho}~,}
where~$\omega_{\mu\nu\rho}$ is the spin connection associated with the Levi-Civita connection.}
\eqn\cont{\eqalign{& {{\, \hat \Gamma}^\mu}_{\nu\rho} = {{\Gamma}^\mu}_{\nu\rho} + {K^\mu}_{\nu\rho}~, \cr
& {{\Gamma}^\mu}_{\nu\rho} = \half g^{\mu \lambda} \left(\d_\nu g_{\rho\lambda} + \d_\rho g_{\nu\lambda} - \d_\lambda g_{\nu\rho}\right)~, \cr
 & K_{\mu\nu\rho} = - K_{\rho\nu\mu}~.}}
Here~${K^\mu}_{\nu\rho}$ is the contorsion tensor, which vanishes for the Levi-Civita connection. It is related to the torsion tensor~${T^\mu}_{\nu\rho}$ as follows,\foot{Given any one-form~$X_\mu$, the torsion tensor~${T^\mu}_{\nu\rho}$ satisfies the defining relation
$$
\hat \grad_\mu X_\nu-\hat \grad_\nu X_\mu=\d_\mu X_\nu-\d_\nu X_\mu -{T^\lambda}_{\mu\nu} X_\lambda~.
$$
}
\eqn\conttor{{T^\mu}_{\nu\rho}={K^\mu}_{\nu\rho}-{K^\mu}_{\rho\nu}~,\qquad K_{\mu\nu\rho}=\ha \left(T_{\mu\nu\rho}+T_{\rho \mu\nu }-T_{\nu\rho\mu}\right)~. }

If we also demand that~$\hat \grad_\mu$ is compatible with the complex structure, i.e.~$\hat \grad_\mu {J^\nu}_\rho = 0$, we find a one-parameter family of allowed connections, parametrized by the following choice of contorsion,
\eqn\conns{K_{\mu\nu\rho} ={1 - t \over 2} {J_\nu}^\lambda \left(dJ\right)_{\lambda\mu\rho} + {t \over 2} {J_\mu}^\alpha {J_\nu}^\beta {J_\rho}^\gamma \left(dJ\right)_{\alpha\beta\gamma}~, \qquad t \in \R~,}
where~$\left(dJ\right)_{\mu\nu\rho} = \grad_\mu J_{\nu\rho} + \grad_\nu J_{\rho\mu} + \grad_\rho J_{\mu\nu}$. To verify that~$\hat \grad_\mu {J^\nu}_\rho = 0$, we use the identity
\eqn\intjid{2 \grad_\mu J_{\nu\rho} + {J_\nu}^\alpha {J_\rho}^\beta \left(dJ\right)_{\mu\alpha\beta} -\left(dJ\right)_{\mu\nu\rho} = 0~,}
which follows from the integrability of~${J^\mu}_\nu$, i.e.~the vanishing of its Nijenhuis tensor.\foot{The Nijenhuis tensor of~${J^\mu}_\nu$ is defined by
$${N^\mu}_{\nu\rho} = {J^\lambda}_\nu \grad_\lambda {J^\mu}_\rho - {J^\lambda}_\rho \grad_\lambda {J^\mu}_\nu -{J^\mu}_\lambda \grad_\nu {J^\lambda}_\rho + {J^\mu}_\lambda \grad_\rho {J^\lambda}_\nu~.$$}

Some connections in the family~\conns\ parametrized by~$t$ have various other desirable properties. We will use the Chern connection~$\grad_\mu^c$, which corresponds to~$t=0$. It has the following useful properties:
\item{1.)} The torsion~${\left(T^c\right)^\mu}_{\nu\rho}$ of the Chern connection satisfies
\eqn\tcprop{\left({\delta_\alpha}^\nu + i {J_\alpha}^\nu\right) \left({\delta_\beta}^\rho - i {J_\beta}^\rho\right) {\left(T^c\right)^\mu}_{\nu\rho} = 0 \quad \Longleftrightarrow \quad {\left(T^c\right)^\mu}_{i \b j} = 0~,}
where~$i, j$ are holomorphic indices with respect to~${J^\mu}_\nu$. We will also need the fully holomorphic components,
\eqn\tchol{{T^k}_{ij} = i {\left(dJ\right)^k}_{ij}~.}
\item{2.)} The Chern connection acts simply on sections~$\omega$ of the canonical bundle~$\CK$ of complex~$(2,0)$-forms, 
\eqn\cdb{\grad^c_i \omega = \Big(\d_i - \half \d_i \log g \Big) \omega~, \qquad \grad^c_{\b i} \omega = \d_{\b i} \omega~, \qquad g = \det(g_{\mu\nu})~.}

\listrefs

\end